\documentclass[sigconf]{acmart}

\usepackage{float}
\usepackage{placeins}

\setcopyright{acmlicensed}
\copyrightyear{2027}
\acmYear{2027}
\acmDOI{XXXXXXX.XXXXXXX}

\begin{document}

\title{Feeling \textsc{Wistful}: Reflecting on Scholarly Sensibilities with Creative Reading Traces}


\author{Sophia W. Liu} 
\affiliation{%
  \institution{UC Berkeley}
  \city{Berkeley}
  \state{California}
  \country{USA}}
\email{sophiawliu@berkeley.edu}

\author{Kate Chier}
\affiliation{%
  \institution{UC Berkeley}
  \city{Berkeley}
  \state{California}
  \country{USA}}
\email{katechier@berkeley.edu}

\author{Shm Garanganao Almeda}
\affiliation{%
  \institution{UC Berkeley}
  \city{Berkeley}
  \state{California}
  \country{USA}}
\email{shm.almeda@berkeley.edu}

\author{Max Kreminski}
\affiliation{%
  \institution{Cornell Tech}
  \city{New York}
  \state{New York}
  \country{USA}}
\email{mkremins@cornell.edu}

\author{Bjoern Hartmann}
\affiliation{%
  \institution{UC Berkeley}
  \city{Berkeley}
  \state{California}
  \country{USA}}
\email{bjoern@berkeley.edu}


\begin{abstract}
Researchers often read before they can articulate what they are looking for. As AI increasingly mediates scholarly search and synthesis, understanding and preserving the idiosyncratic judgments guiding early exploration become important. We call these evolving orientations \textit{scholarly sensibilities}. To understand curiosity-driven reading, we first examined Wikipedia rabbit holing, a self-directed browsing practice, then designed \textsc{Wistful}, a research probe for open-ended scholarly exploration that captures reading paths as \textit{creative reading traces}. We studied \textsc{Wistful} with 16 HCI researchers---eight junior and eight senior---in comparison with their usual workflows. Researchers approached the same scholarly landscape differently, with familiarity and personal interests shaping what they pursued and semantic proximity and scholarly links shaping their paths. Their traces made these differences visible and let readers revisit and compare their paths. We position creative reading traces as artifacts for reflection and exchange and as a means of studying how scholarly sensibilities are expressed through reading.
\end{abstract}



\begin{teaserfigure}
 \includegraphics[width=\textwidth]{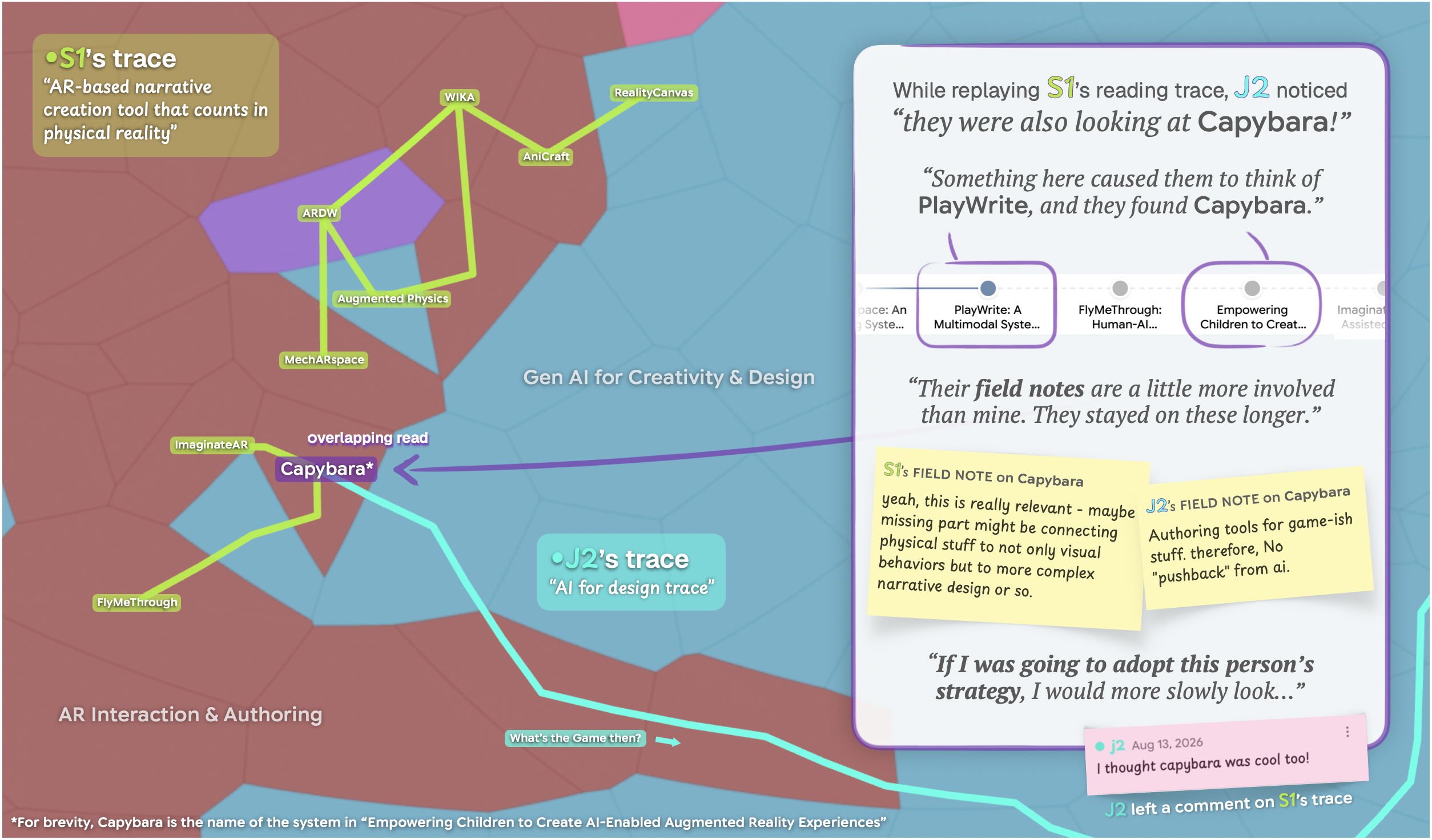}
  \caption{A junior and senior researcher reached the same paper through different reading paths. Left: J2's and S1's traces converge on \textsc{Capybara} near the GenAI Creativity--AR boundary, with J2 moving from familiar GenAI work into AR and S1 exploring AR from its boundary with familiar GenAI work. Right: while replaying S1's trace, J2 noticed their overlap and contrasting Field Notes and considered adopting parts of S1's reading strategy.}
  \Description{description if any}
  \label{fig:teaser}
\end{teaserfigure}

\maketitle


\section{Introduction}

Researchers often read before they know exactly what they are looking for. As a research direction takes shape, they follow papers and connections that catch their attention, drawing on prior knowledge and personal experience to decide what seems worth pursuing. These judgments reflect what we call \textbf{\textit{scholarly sensibilities}}, evolving orientations that guide what researchers notice and pursue in the literature~\cite{Sensibilities, professionalVision, aQuestionOfTaste}. Because these orientations can guide attention before they are fully articulated, they can be expressed in the idiosyncratic paths researchers take during exploratory reading.

AI increasingly mediates these early encounters with literature as researchers now use LLMs for information seeking and synthesis. Recent work suggests that such mediation can concentrate what researchers read and ultimately cite~\cite{citationMonoculture} and lead different users toward more similar ideas~\cite{homogenizationEffects}. These effects may be especially consequential before a research direction is fully formed, when researchers rely on orientations they cannot yet clearly articulate as a query or prompt to decide what seems worth pursuing. Understanding how these scholarly sensibilities are expressed through reading, and how they develop through repeated encounters with literature, therefore provides a basis for studying how AI-mediated tools may shape them over time.

One way to study these orientations is to preserve the paths through which they are expressed. Building on \textit{creative reading} as an expressive process in which readers construct their own readings through what they notice, follow, and connect~\cite{CreativeReading}, we use the term \textbf{\textit{creative reading traces}} for partial representations of this constructive process. These traces preserve the path a reader makes through literature for later revisitation and interpretation~\cite{TracingCreativity, RATs}. By making these otherwise ephemeral expressions visible, creative reading traces provide material for studying scholarly sensibilities.

To understand open-ended reading when exploration is primarily driven by curiosity rather than a predetermined information need, we first studied Wikipedia ``rabbit holing,'' a self-directed browsing practice often pursued for its own sake~\cite{goingDownRabbithole, whyWikipediaBeyondEnglish, dialecticalActivities}. These observations informed the design of \textsc{Wistful}, a research probe that brings qualities of this curiosity-led exploration into scholarly reading through an open spatial landscape where researchers can follow what catches their attention without first articulating a precise query. As readers explore, \textsc{Wistful} preserves their paths and annotations as creative reading traces that can be revisited and shared.

We use \textsc{Wistful} to investigate how scholarly sensibilities are expressed through open-ended reading and how preserving those reading paths can make aspects of these sensibilities visible to readers, to one another, and to us as researchers. We ask: (1) How are researchers' scholarly sensibilities expressed during exploration of a shared literature? (2) What becomes possible when these explorations are preserved as creative reading traces for reflection and exchange?

In this work, we contribute \textsc{Wistful}, a research probe that lays out an open scholarly landscape in which readers can follow emerging interests before they can fully articulate what they are seeking, while preserving the paths they take as creative reading traces. We studied \textsc{Wistful} with 16 HCI researchers---eight junior and eight senior---in comparison with their usual scholarly reading workflows. We find that researchers read the same landscape through different personal reference frames, with accumulated experience and personal interests shaping what they pursued. Their traces revealed how these orientations shaped movement through semantic and scholarly structures as inquiries developed. Revisiting and sharing those traces also made readers' paths available for reflection and comparison, while giving us empirical material for examining how scholarly sensibilities were expressed through reading. These findings position creative reading traces as a way to study expressions of scholarly sensibility and suggest how reading systems might preserve space for idiosyncratic exploration as AI increasingly mediates scholarly reading.

\section{Related Work}

Our work draws together three strands of prior research. We first situate scholarly sensibilities in work on professional vision, scientific taste, and creative reading. We then consider scholarly exploration as movement through a mediated information landscape, including the growing role of generative AI in shaping what readers encounter and pursue. Finally, we draw on work on interaction histories and reflective practice to motivate \textit{creative reading traces} as artifacts that make reading processes available for reflection and study.

\subsection{Developing Scholarly Ways of Seeing}

Expertise includes tacit ways of knowing developed through practice and participation in a community~\cite{personalKnowledge, tacitDimension, TEASet}. Goodwin's \textit{professional vision} describes how members of a profession learn to see and interpret phenomena in their domain~\cite{professionalVision}, including through collective reading and critique in science~\cite{scientistsProfessionalVision}. Related accounts of designerly ways of knowing and \textit{sensibilities} describe expertise as a cultivated way of perceiving and responding to the world~\cite{designerlyWaysOfKnowing, Sensibilities}. Scientific taste captures one expression of such orientation in judgments about ``what science to do, what not to do,'' shaped through experience and social context~\cite{scientificChoice, aQuestionOfTaste, researchTopicChoice}. Building on these accounts, we use \textbf{\textit{scholarly sensibilities}} to describe researchers' evolving orientations toward what they notice and judge worth pursuing. These sensibilities can be expressed in the problems researchers choose as well as in how they approach and interpret prior work.

Research experience also shapes how researchers enter and navigate an emerging problem space. Prior work finds that less-experienced researchers may fix an initial framing too early or struggle to articulate useful search terms, while more experienced researchers draw more flexibly on prior knowledge to reformulate queries and judge relevance~\cite{revisionStudentvsExpert, bernsteinVectoringResearch, onlineLitSearchComparison}. Research needs can also precede the vocabulary required to express them~\cite{searchStrategiesAcademicLifecycle, queryFormulation}. These studies suggest that experience affects how readily researchers can translate their emerging judgments about what seems relevant or promising into a search. Scholarly sensibilities may therefore guide which possibilities feel worth pursuing before researchers can articulate those judgments as a prompt or query~\cite{gulfEnvisioning}.


Reading is one place scholarly sensibilities are put to work. Reader-centered accounts treat meaning as something made in the encounter between reader and text rather than fixed in advance~\cite{rosenblattTransactional, deathOfTheAuthor}. \textit{Creative reading} extends this view to scholarly reading, where readers create interpretations while developing the orientations they bring to future texts~\cite{CreativeReading}. Creative experience locates such change in open-ended person-world encounters, while dialectical activity emphasizes how repeated engagement in research can deepen one's understanding of the intrinsic value of the activity~\cite{CreativeExperience, dialecticalActivities}. Reading therefore offers a site where scholarly sensibilities are both expressed and cultivated. A broader account of scholarly judgment becomes especially important as interfaces, including generative AI, increasingly mediate what researchers encounter and how they interpret prior work.

\subsection{Scholarly Exploration and Mediated Attention}

Open-ended literature exploration rarely follows a predetermined path. Information seeking can unfold through successive discoveries that reshape what a researcher notices and interprets~\cite{informationSearchProcess, informationForaging, berrypicking, exploratorySearch, orienteering}. The \textit{information flâneur} frames such exploration as movement between broad exploration and deeper immersion as interests develop~\cite{informationFlaneur}. Debord's \textit{dérive} similarly describes drifting through a terrain according to both personal attractions and the contours of the environment~\cite{Debord}. Gamboa and Andersen connect the \textit{dérive} to the formation of sensibilities, describing an ongoing practice of wandering while remaining attentive and alert to what emerges~\cite{Sensibilities}. Such openness can be generative in research, where serendipitous encounters, distant analogies, and atypical combinations can surface connections that are difficult to anticipate in advance~\cite{serendipityKnowledgeWork, augmentingSciCreativityAnalogicalSearch, atypicalCombinations}.

This motivates our conception of \textit{scholarly dérive}, where movement through literature is guided by a researcher's evolving sensibilities while the encounters along that path can, in turn, reshape them. Scholarly dérive can therefore produce both a distinct reading and, through this dialectical process, a changed reader~\cite{CreativeReading, dialecticalActivities}.

Scholarly dérive emerges through an interaction between the reader and the terrain through which they move~\cite{Debord, informationCity}. Researchers approach a literature differently depending on their familiarity and experience~\cite{infoSeekingBehaviorCS, keepingUptoDate, onlineLitSearchComparison, searchStrategiesAcademicLifecycle}. Interfaces form part of this terrain by giving structure to the relationships a reader can encounter and follow. Hypertext and spatial systems make such relationships navigable through links and visual organization~\cite{tedHypertext, spatialHypertextInfoTriage, informationCity}, while scholarly reading systems extend these structures to research literature~\cite{SemanticReaderProject, AbstractExplorer, CiteSee, Threddy, gardenOfPapers}. Such representations shape the conditions of exploration, while readers construct meaning through how they navigate and relate what they encounter~\cite{Artographer, algorithmicWaysofSeeing, hypertextualFriction, CreativeReading}.

Generative AI increasingly mediates this terrain, synthesizing and framing information in ways that can reduce engagement with underlying sources~\cite{Liao2024LLMsAR, pewAISummaries2025}. These systems now participate in exploratory search and early-stage ideation, with their influence varying by research experience~\cite{AIinEarlyStageResearch, whoUsesAIinResearch, exploratorySearchGenAI, experienceImpactAISuggestionsPercetionsIdeas}. Less-experienced researchers may incorporate AI suggestions more readily, and AI can become an intermediate layer before consultation with human experts~\cite{experienceImpactAISuggestionsPercetionsIdeas}. This mediation redistributes cognitive and epistemic work, shaping which evidence receives attention and which directions appear worth pursuing~\cite{cognitiveOffloading, toTrustorThink, epistemicAgencyAgeofLLMs}.

Concerns about homogenization in AI-supported ideation and growing citation monocultures suggest that AI mediation may lead different researchers toward more similar ideas and sources~\cite{homogenizationEffects, citationMonoculture}. This raises a design challenge of helping researchers articulate emerging interests without steering them toward model-favored directions~\cite{dearthAuthor}. Recent work has also begun to computationally model scientific taste from traces of research activity~\cite{AIScientificTaste}. Yet scholarly sensibilities develop through prolonged, embodied, and varied experience~\cite{Sensibilities, CreativeReading}. AI mediation therefore changes the conditions under which researchers exercise and develop these ways of seeing. The paths through which researchers read and choose what to follow become meaningful \textit{traces} of how those sensibilities are expressed.

\subsection{Trails and Traces for Reflection}

The significance of traces follows from a broader shift in creativity research toward the process through which creative work transpires~\cite{ProcessDefinition, BeyondProductivity}. Recent work on \textit{creative activity traces} treats iterations, references, and other process data as meaningful records of creative activity~\cite{TracingCreativity}. If scholarly reading is itself a creative activity~\cite{CreativeReading}, this motivates \textbf{\textit{creative reading traces}}: partial records of the choices through which an exploratory reading path takes shape~\cite{CreativeReading, RATs}.


Interaction histories provide a rich precedent for making exploration traceable. Bush's foundational \textsc{Memex} imagined associative trails that could be stored and revisited~\cite{AsWeMayThink}. As hypertext made linked navigation commonplace~\cite{tedHypertext}, systems such as \textsc{Read Wear} and \textsc{Footprints} made these histories visible as paths through documents~\cite{editWearAndReadWear, Footprints}. Scholarly search systems have likewise preserved activity across a literature search~\cite{searchActivityTraces}. Research on browsing history and personal informatics further shows how records of prior activity can support recollection and reflection~\cite{browsingHistoryVis, reflectionOnInfoGatheringBrowsingHistory, personalInformatics}, including the reflection-on-action Schön describes as making sense of past experience~\cite{reflectivePractitioner}.

Shared search histories, notes, and collective sensemaking structures can expose others' paths through information~\cite{SearchTogether, notesToSelfToShare, Semble}, reflecting how scholarly ways of seeing are socially cultivated~\cite{professionalVision, Sensibilities}. Engelbart envisioned augmentation across horizons from ``twenty minutes'' to ``twenty years'' as part of an integrated way of working rather than merely support for isolated tasks~\cite{augmentingIntellect}. Creative reading extends this long-horizon view by centering the transformation of the reader through reading and the sensibilities they carry into future texts~\cite{CreativeReading}. Creative reading traces preserve fleeting expressions of this process, making aspects of a reader's developing sensibilities available for reflection and growth.

\section{Formative Study}

We studied wiki ``rabbit holing'' on Wikipedia as a form of self-directed, curiosity-driven exploration we hoped to support in scholarly reading. Named after Alice following the White Rabbit, wiki rabbit holing describes following one's curiosity from topic to topic through linked information~\cite{wikiRabbitHole}. Intrinsic learning---reading to satisfy personal curiosity---is the most commonly reported motivation for reading Wikipedia~\cite{whyWikipediaBeyondEnglish}. Prior work has characterized Wikipedia exploration at scale through surveys, navigation logs, and the network structure of readers' paths~\cite{whyWikipedia, whyWikipediaBeyondEnglish, goingDownRabbithole, wikipediaCuriosityStyles}. We instead conducted an observational qualitative study of rabbit holing to understand individual readers' aesthetic experiences of curiosity-driven exploration.

We conducted sessions with $N=9$ participants who varied in background, Wikipedia use, and rabbit-holing frequency (Appendix Table~\ref{tab:formative-participants}). Starting articles were selected based on participants' stated interests. 


\subsection{Procedure}

Before each session, participants completed a screener about their Wikipedia use, preferred learning sources, and three topics of curiosity. We selected starting articles related to these interests, varying participants' familiarity and relationships to the topics and the kinds of topics explored. Participants then browsed Wikipedia for 15 minutes while thinking aloud, freely following links, searching, and backtracking. Immediately afterward, they closed Wikipedia and reconstructed their exploration from memory, describing what they encountered, why they moved between topics, and what stood out or felt meaningful.

We then conducted a semi-structured interview about participants' prior knowledge, relationships to encountered topics, and typical information-seeking and Wikipedia browsing practices.


Informed by constructivist grounded theory~\cite{groundedTheory}, we used open coding~\cite{codingManual} to analyze navigation traces, think-aloud data, and interview transcripts. Three authors participated in coding, with two authors coding each participant. Emerging themes guided theoretical sampling of subsequent participants and starting topics, and were iteratively refined through discussion into the findings below.


\subsection{Key Formative Study Findings}

By tracing how participants followed their curiosity through Wikipedia, we identified what shaped the directions they pursued and what remained meaningful afterward. Participants visited 2--21 unique articles during the 15-minute browse ($Mdn = 15$), with paths ranging from mostly linear sequences to branching and looping explorations.

\subsubsection{Inquiry Emerged Through the Path}
Participants often developed questions from what they encountered en route. P2 described \textit{``loose threads in my brain''} that became apparent after \textit{``seeing words on the page that resonated,''} comparing shifts in direction to how \textit{``one push of wind can push you one direction or another.''} Questions also grew from prior knowledge. P1 described questions emerging from vague memories, finding that \textit{``in trying to answer those questions, I'll come up with new questions.''} Browsing also resurfaced old questions participants had not pursued before, as when P8 was reminded of questions they had never made time to investigate. Paths could branch into temporary inquiries before returning to an earlier thread. P6 began with the history of \texttt{Monte Carlo method}, followed its connection to nuclear weapons research into Manhattan Project history, then \textit{``wrapped up that rabbit hole''} and returned.

\subsubsection{Curiosity Reflected Personal Ways of Seeing}
What participants noticed and pursued reflected what they knew, cared about, and identified with. P4 described familiar topics as \textit{``comfort subject[s]''} and a tendency to \textit{``focus on the people and their lives.''} Encountering a scandal from their home country prompted, \textit{``how did I not know about this?''} By contrast, P5 deliberately skipped familiar topics around the \texttt{Whole Earth Catalog} in search of \textit{``more unknown territories.''} P7 opened \texttt{Joni Mitchell} after recognizing an artist whose song they had recently sung in choir, while P6's nuclear engineering background shaped which mathematical links they pursued. Curiosity also reflected who participants wanted to become. P2 described their topics as things \textit{``the ideal [P2] would know a lot about.''}

\subsubsection{Readers Reconstructed Both Paths and Moments}
When asked to recount their browsing, participants emphasized both the broader shape of their paths and specific moments. P1 said seeing their tabs or history would help them remember where they had gone, yet recalled one part of the browse as \textit{``inspirational''} for a story they wanted to write. P5 described two different paths from their starting article and surprising connections between them. P7 could not remember how they reached \texttt{Kanye West}, but recalled becoming curious about his controversies. When asked to visualize their browsing, participants described trees with branches and graphs with nodes and edges. They also imagined related topics spatially, as nested or overlapping circles and clusters arranged across two or three dimensions. These different ways of recalling and representing a browse motivated supporting reflection through multiple views of the same navigation history.

\section{\textsc{Wistful}}


Drawing on our formative study and prior work, \textsc{Wistful} is a research probe for open-ended scholarly exploration through the kind of self-directed reading we describe as \textit{scholarly dérive}. Our formative study showed how curiosity-driven browsing develops through readers' questions, prior knowledge, and personal interests, and how readers later reconstruct these explorations through both specific moments and broader path structures. These findings informed the design of \textsc{Wistful} as an environment for exploring scholarly literature, capturing reading paths as traces, and revisiting those traces during and after exploration.

\subsection{Design Goals}
We synthesized these insights into three design goals for \textsc{Wistful}.

\begin{description}

\item[\textbf{DG1:}]
\textbf{Support self-directed, curiosity-driven exploration.} Open-ended exploration can surface questions and possibilities readers would not know to seek in advance~\cite{onlineLitSearchComparison, informationFlaneur, Artographer}, a pattern we also observed in our formative study. \textsc{Wistful} should let readers follow what catches their interest as their inquiry develops.

\item[\textbf{DG2:}]
\textbf{Capture what readers notice and pursue.} Our formative study showed that reading paths reflected participants' familiarity, personal interests, and backgrounds, aligning with prior work on sensibilities and scientific taste~\cite{Sensibilities, scientificChoice}. \textsc{Wistful} should record these paths and provide lightweight ways for readers to capture what draws their attention and why.

\item[\textbf{DG3:}]
\textbf{Support reflection on and exchange of reading traces.} Participants reconstructed their explorations through both specific moments and broader path structures. Creative activity traces can make ephemeral processes available for reflection~\cite{TracingCreativity}, while shared histories can expose others' paths and associations~\cite{SearchTogether, notesToSelfToShare}. \textsc{Wistful} should present reading traces through multiple representations that support retrospective interpretation and comparison with other readers.

\end{description}



\subsection{System Overview}
\begin{figure*}
\centering\includegraphics[width=0.99\linewidth]{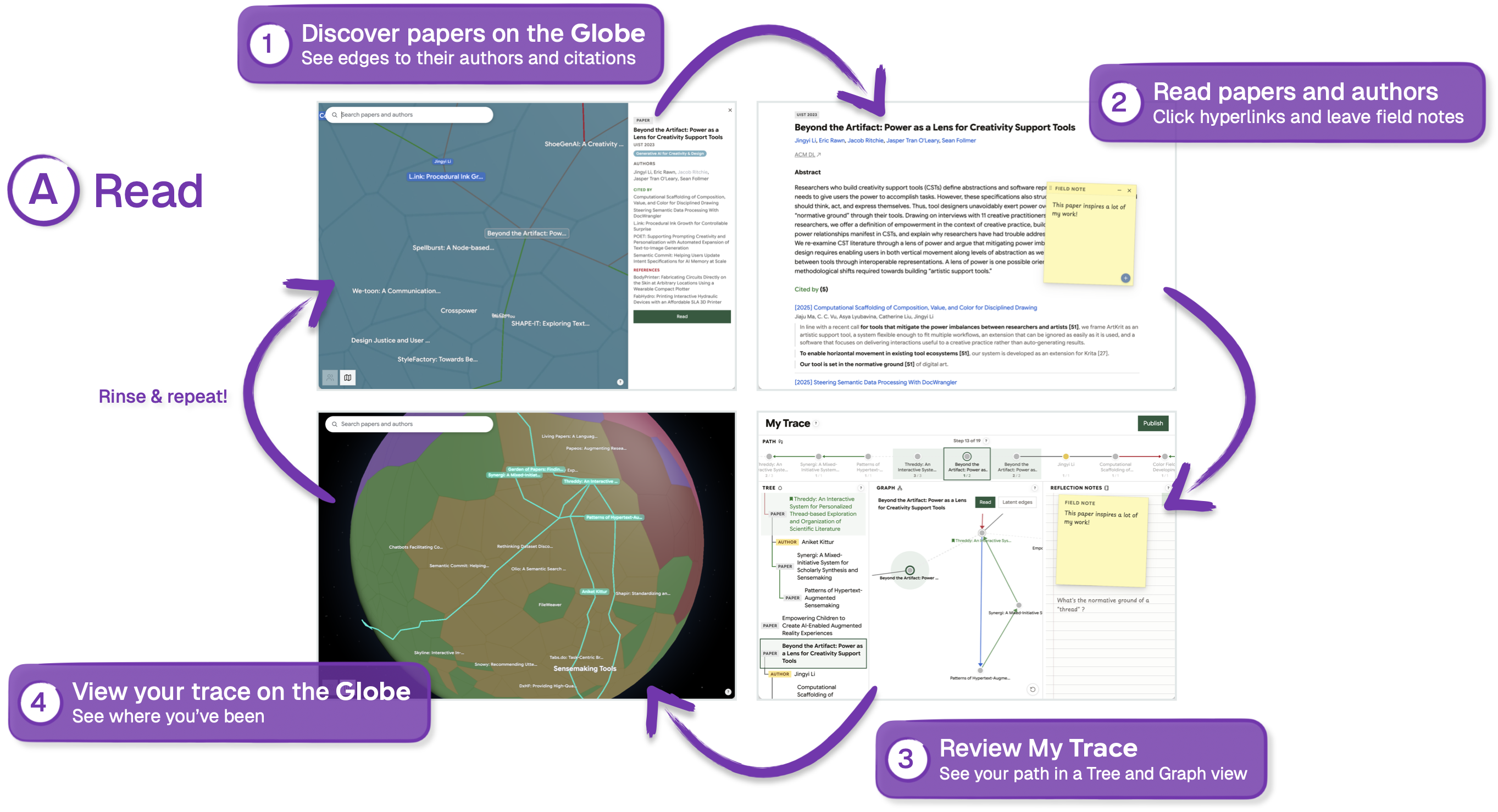}
    \caption{Reading workflow in \textsc{Wistful}. Readers (1) discover papers and scholarly links on the Globe or Map, (2) read papers and authors while leaving Field Notes and following hyperlinks, (3) review their accumulated path in My Trace, and (4) return to the landscape to see where they have been and continue exploring.}
    \Description{Workflow diagram for Wistful’s Read phase. Four interface screenshots are arranged in a cycle: 1) discover papers on the Globe and follow author or citation edges; 2) read papers and authors, follow hyperlinks, and leave field notes; 3) review the resulting reading path in My Trace using tree and graph views; and 4) return to the Globe to see the spatial trace of where you have been. Purple arrows connect the four steps in a repeating loop labeled “Rinse \& repeat!”}
    \label{fig:workflow_1}
\end{figure*}
\textsc{Wistful} builds on a previously demonstrated system for spatial scholarly exploration and reading traces~\cite{UISTfulDemo}. The name reflects looking back over a reading path and the longer personal histories through which scholarly sensibilities develop. It presents 771 UIST papers from 2020--2025 as a shared scholarly landscape and records readers' navigation through this corpus as creative reading traces. Its three main interfaces are the \textit{Globe} and corresponding 2D \textit{Map}, for exploring and reading; \textit{My Trace}, for revisiting and annotating one's reading trace; and the \textit{Trace Library}, for publishing traces that others can replay, compare, and discuss. As readers move between papers and authors, \textsc{Wistful} records the ordered sequence of visits and transitions, together with notes made during reading. This trace remains private until the reader chooses to publish it.






\subsection{Constructing the Scholarly Landscape}

\textsc{Wistful} includes 771 full papers from the UIST main proceedings from 2020--2025, together with their authors and citation relationships. We chose UIST because it offers a bounded interactive-systems community spanning diverse research areas, while its frequent use of short system and project names makes papers identifiable as map labels. We use this corpus as a common space for exploration rather than as a representation of HCI as a whole.

We embed each paper's title and abstract using \texttt{all-MiniLM-L6-v2}, producing a 384-dimensional semantic representation, and project the embeddings onto a sphere using spherical UMAP. Papers with more similar embeddings are positioned closer together. We use a sphere to present the bounded corpus as a continuously navigable whole without planar edges, following prior work on spherical information visualization~\cite{sphereViz}. The precomputed layout is shared across readers and can be explored as either a 3D Globe or a 2D equirectangular Map. This representation reflects how participants in our formative study described related topics as clusters and regions across two or three dimensions and draws on prior spatial interfaces including \textsc{Paperscape} and \textsc{Artographer}~\cite{Paperscape, Artographer}. We treat it as one computational organization of the corpus rather than a definitive organization of the field.

We further divide the semantic embeddings into seven ``continents'' using \(k\)-means, choosing \(k=7\) to extend the globe metaphor. \texttt{GPT-4o-mini} labels each cluster with a topical description used throughout the system and study. These continents provide coarse landmarks for navigation. Each paper occupies a spherical Voronoi cell, while authors with at least two papers are placed at the spherical mean of their papers, yielding 635 author landmarks. All 2,737 authors remain accessible through paper and author pages.

In-corpus citation and authorship relationships provide another structure through the same corpus. These links can connect papers that are distant in the semantic layout, giving readers routes across regions that spatial proximity alone would not expose. To display these relationships across the globe, \textsc{Wistful} routes each link along the shortest path through a spherical Delaunay graph connecting neighboring paper and author sites. The Delaunay graph determines how links are drawn, while the relationships themselves come from the scholarly corpus.






\subsection{Exploring and Reading the Landscape}

Readers can move freely between the 3D Globe and 2D Map, which provide equivalent views of the same scholarly landscape. Semantic zoom progressively reveals continent, paper, and author labels as readers move closer to a region. Selecting a paper or author reveals information about that node and its relationships.

As a research probe for scholarly dérive, \textsc{Wistful} prioritizes exploratory movement and emerging interests over optimized retrieval. Search therefore provides lightweight lookup by paper title or author, while spatial proximity and scholarly links remain the primary ways of discovering new work. Paper titles are truncated to reduce visual clutter. Because UIST titles often begin with a short system or project name, this name typically remains visible, with the full title available upon selection.

Opening a paper shows its title, authors, abstract, and the papers it cites and that cite it within the corpus, with the full ACM reference list available separately. Author pages show the papers an author has contributed to and their co-authors. Readers can follow these relationships or return to the Globe or Map at any time, allowing paths to develop through both semantic proximity and scholarly links.

\subsection{Capturing Creative Reading Traces}

\textsc{Wistful} represents each reading trace in two complementary forms: a temporal representation of the reader's ordered visits and a spatial representation over the scholarly landscape. The temporal representation records visits in sequence, preserving revisits as distinct steps.

Readers can attach \textit{Field Notes} to individual visits through a prominent draggable sticky beside the abstract that asks, ``Why are you here? Where did you come from? Where are you going?'' Placing this prompt directly in the reading interface encourages readers to record what draws their attention and where they might go next. Notes remain attached to a specific visit, so returning to the same page creates a new opportunity to record a response. Readers can also choose to add off-corpus pages to the temporal trace.

The spatial representation builds up as readers explore. Consecutive visits to mapped papers and authors are routed across the Globe and Map, allowing readers to see the path they have accumulated when they return to either view. Visit order determines which hops are drawn, while the spherical Delaunay graph determines how each hop is routed across the surface. Off-corpus pages and unmapped authors remain in the temporal representation without appearing as spatial arcs. Readers choose a color for their trace, which persists across the landscape as they continue reading.


\subsection{Reflecting on Creative Reading Traces}

\textit{My Trace} lets readers revisit their temporal trace through coordinated Path, Tree, and Graph views, reflecting the different path- and structure-based ways participants in our formative study reconstructed past browsing. The Path preserves visits in order, including revisits and external steps. The Tree reveals branching structure by nesting each page beneath the page whose scholarly link led the reader to it. The Graph shows visited nodes together with latent corpus relationships and interpretive links created by the reader.

Readers can scrub through the Path to revisit individual moments. Field Notes written during reading appear read-only beside a separate \textit{Reflection Note} for retrospective interpretation. Reflection Notes can connect visited pages through \texttt{[[wiki]]} links, which appear as blue interpretive edges in the Graph. Readers can also crop a contiguous portion of the trace and add a title and framing note, turning part of a live trace into a reading they may choose to publish.

\subsection{Sharing Reading Traces}
\begin{figure*}
\centering\includegraphics[width=0.99\linewidth]{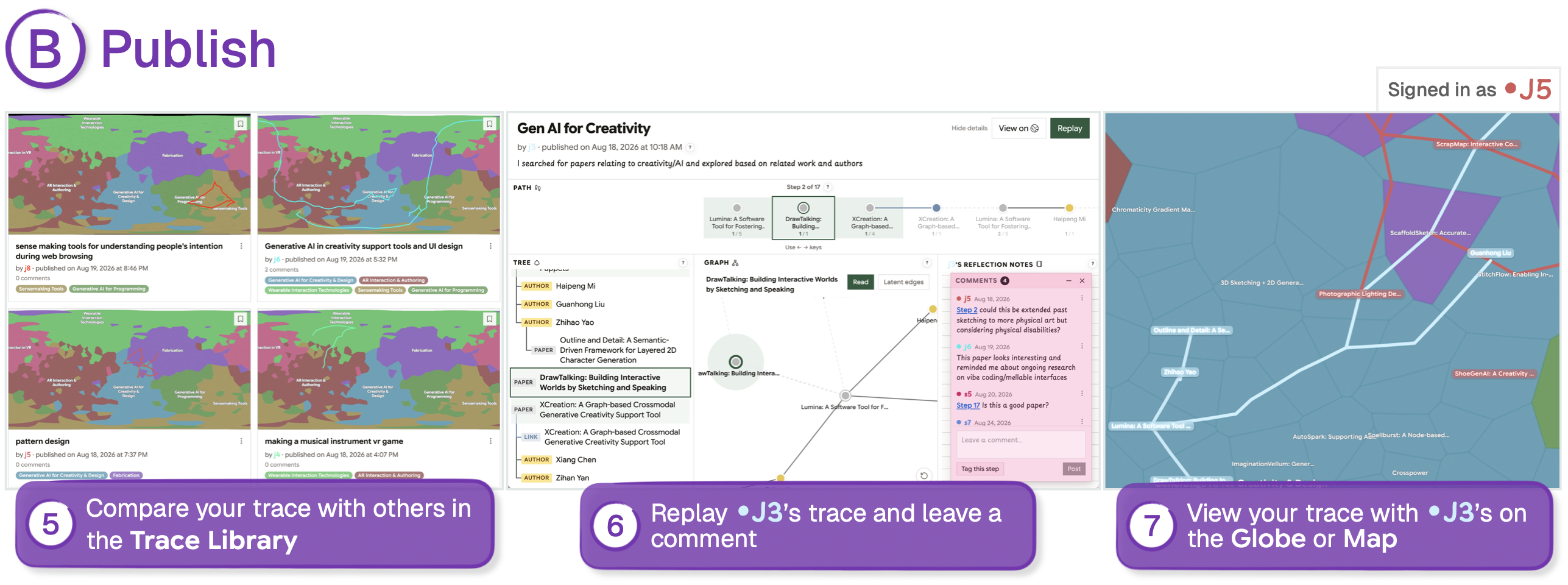}
    \caption{Sharing and publishing workflow in \textsc{Wistful}. Participants could (5) compare traces in the Trace Library, (6) replay another reader's trace and comment on particular moments, and (7) view another trace alongside their own on the Globe or Map.}
    \Description{Workflow diagram for Wistful’s Publish phase. Three interface screenshots show how published traces support comparison and discussion: 5) browse other readers’ traces in the Trace Library; 6) replay J3’s trace in tree and graph views and leave comments on individual steps; and 7) view J5’s trace together with J3’s on the Globe or Map to compare their spatial reading paths.}
    \label{fig:workflow_2}
\end{figure*}
Because \textsc{Wistful} treats reading as a creative activity, the \textit{Trace Library} lets readers share the reading they have created. From \textit{My Trace}, readers select a contiguous portion of their path, give it a title and framing reflection, and choose whether to keep it private or publish it publicly. Publishing saves the selected segment as a fixed snapshot, including its notes and external links, turning a live path into a shareable reading artifact.

Public traces appear in the Trace Library, where others can open and replay them step by step, inspect their notes, or overlay them with their own trace on the Globe or Map (Fig.~\ref{fig:workflow_2}). Readers can also comment on the trace as a whole or attach a comment to a specific step, allowing discussion to refer directly to moments in another reader's path. Published traces can also be aggregated on the Globe or Map, allowing readers' paths to accumulate as visible marks on the shared scholarly landscape~\cite{weWereOnline}.

\subsection{Implementation Details}

\textsc{Wistful} is a React web application. The corpus is constructed from the ACM UIST 2020--2025 main proceedings, with paper metadata and citation relationships enriched through OpenAlex and Crossref. The 3D Globe is rendered with \texttt{react-globe.gl} and Three.js, while the Map uses an equirectangular projection of the same spherical model. Paper embeddings, spherical UMAP coordinates, \(k\)-means continents, Voronoi cells, author positions, and globe textures are computed offline and shipped as static assets, so every reader encounters the same scholarly landscape while reducing runtime computation.

Reading traces are recorded in the client as readers navigate. Scholarly links and reading-trace hops are routed at runtime over the precomputed spherical Delaunay graph using Dijkstra's shortest-path algorithm with geodesic edge lengths. A FastAPI backend maintains an event log for later analysis and stores published traces as fixed snapshots that can be replayed independently of the reader's live browsing state.

\section{Method}

We use \textsc{Wistful} as a research probe, following prior work on technology probes and discovery-oriented systems research~\cite{probeHCI, systemsPaperDead}, to examine how scholarly sensibilities are expressed through open-ended reading among junior and senior HCI researchers. We analyze creative reading traces alongside think-aloud and interview data to examine how prior experience, familiarity, and personal interests shape exploration, focusing on brief \textit{expressions} of existing scholarly sensibilities and what becomes possible when reading paths are made visible for reflection and exchange.

\subsection{Study Design}

We conducted a mixed-methods, within-subjects study with 16 HCI researchers. Each participant completed two 15-minute literature-exploration tasks: one using their usual workflow and one using \textsc{Wistful}. Participants began from an assigned topic and explored toward a possible line of inquiry, while the path and resulting direction remained open. We designed these as bounded elicitation episodes of scholarly dérive rather than approximations of a complete literature review; dérive can unfold over hours or days, but may also occur in ``fairly brief moments''~\cite{Debord}. Think-aloud during exploration and a longer interview afterward provided insight into the judgments underlying participants' choices.

Task framing matched participants' research roles. Junior participants developed a line of inquiry they might pursue for an HCI research-class project, while senior participants identified one they would be interested in pursuing \textit{with} a student. This broad prompt gave participants a comparable reason to engage with the literature without prescribing what they should find or how they should proceed. Participants rated their familiarity with each of \textsc{Wistful}'s seven topic continents before the study. Junior participants received two topics from the more familiar end of their ratings so they had enough grounding to develop a plausible direction within the short task, while senior participants received topics from the less familiar end to reduce reliance on topic-specific expertise and examine how broader research experience shaped exploration in less familiar literature. Within each pair, one topic remained relatively more familiar than the other. Because familiarity and task framing varied with cohort by design, we treat junior--senior differences as descriptive rather than causal. Within each cohort, we used a \(2 \times 2\) counterbalancing design crossing workflow order with relative topic familiarity, with two participants per cohort in each cell.

We used participants' usual literature-search practices as an ecologically grounded reference condition rather than a standardized control. Participants could use the tools and strategies they ordinarily relied on, including AI assistants. This grounded comparison of \textsc{Wistful} in participants' established practices and gave the subsequent interview a concrete basis for reflection.






\subsection{Recruitment \& Participants}
We recruited 16 HCI researchers across career stages through research-community channels, professional networks, and snowball sampling. We classified participants primarily by years of HCI research experience, using experience across multiple projects and mentoring others to resolve ambiguous cases. Eight participants comprised each cohort, denoted J1--J8 and S1--S8.

The junior cohort included master's and early-stage PhD researchers, a postbaccalaureate researcher, an industry researcher, and a teaching faculty member; six had fewer than three years of HCI research experience and two had three to five years. The senior cohort included late-stage PhD researchers, an industry research scientist, and faculty; seven of eight had at least six years of HCI research experience, including three with more than 20 years. All senior participants had mentored junior researchers, supporting the advising role used in their task framing. Appendix Table~\ref{tab:participants} summarizes participants' backgrounds and study assignments.

\subsection{Procedure}

Sessions were conducted over Zoom and lasted approximately 90 minutes. Participants completed both tasks in counterbalanced order while sharing their screen and thinking aloud. A personal Google Doc was available for optional note-taking. After each task, participants stated the line of inquiry that emerged and what led them to it.

\subsubsection{\textsc{Wistful}}

Before the \textsc{Wistful} task, participants watched a short tutorial and practiced with the system before the moderator reset their trace. During the task, participants were encouraged to use Field Notes and reminded that My Trace was available for review. At the end, they published their trace to the Trace Library.

Participants then replayed other published traces and commented on one. Because traces accumulated over the course of the study, participants encountered different numbers of prior traces. The first two participants were seniors, followed by all eight juniors and the remaining six seniors. J1--J3 were directed to senior traces; from J4 onward, participants could choose any trace and were told whether its author was a junior or senior researcher.

\subsubsection{Usual Workflow}

For the usual-workflow task, participants used their usual literature-discovery tools and strategies, including scholarly search engines, web search, and AI assistants. The moderator reminded them of the practices they had reported in the screener but otherwise imposed no restrictions.

\subsubsection{Post-Task Questionnaire and Interview}

After each task, participants completed a short questionnaire about their exploration experience and resulting line of inquiry. The full questionnaire is provided in Appendix~\ref{app:post-task-questionnaire}. After both tasks, we conducted a semi-structured interview comparing the workflows and probing the experiences and judgments shaping participants' exploration. Senior interviews additionally addressed mentorship and the development of scholarly sensibilities.

\subsection{Data Collection and Analysis}
We recorded participants' screens, think-aloud, and interviews, and collected their shared notes, stated lines of inquiry, survey responses, and screener data. In \textsc{Wistful}, we additionally collected interaction logs, Field and Reflection Notes, published traces, and Trace Library comments. Three authors analyzed these materials together, open-coding transcripts alongside traces, notes, and logs~\cite{codingManual}. We used logged behavior to establish what participants did and participants' accounts to interpret the experiences and judgments underlying those actions. Beginning with a working vocabulary around familiarity, traces, notes, and spatial versus hyperlink movement, we refined themes through discussion and synthesis across participants.

We derived descriptive measures from interaction logs to characterize exploration. For search, we collapsed repeated and prefix queries and computed Shannon entropy over distinct queries. To characterize movement through the landscape, we measured geodesic distance and continent crossings between consecutive paper first-opens, distinguished map and hyperlink transitions, and computed continent entropy over opened papers. We related these paths to participants' familiarity ratings by comparing the most-visited non-assigned continent with their highest-familiarity alternative and used a local boundary analysis for the AR--GenAI cases reported below. We segmented Google Docs and Field Notes by line and classified segments as links, references, short labels, quotations, or participants' own wording. Questionnaire responses and trace-use measures were analyzed descriptively alongside the qualitative data.

\section{Findings}

Our findings show how scholarly sensibilities were expressed through open-ended exploration. Participants brought personal reference frames to the shared scholarly map, followed familiarity and personal interests across its boundaries, and composed reading paths across semantic and citation structures. Preserving these paths as traces supported orientation during exploration, made differences between readers visible, and created opportunities for later reflection and exchange.

\begin{figure}
\centering\includegraphics[width=0.99\linewidth]{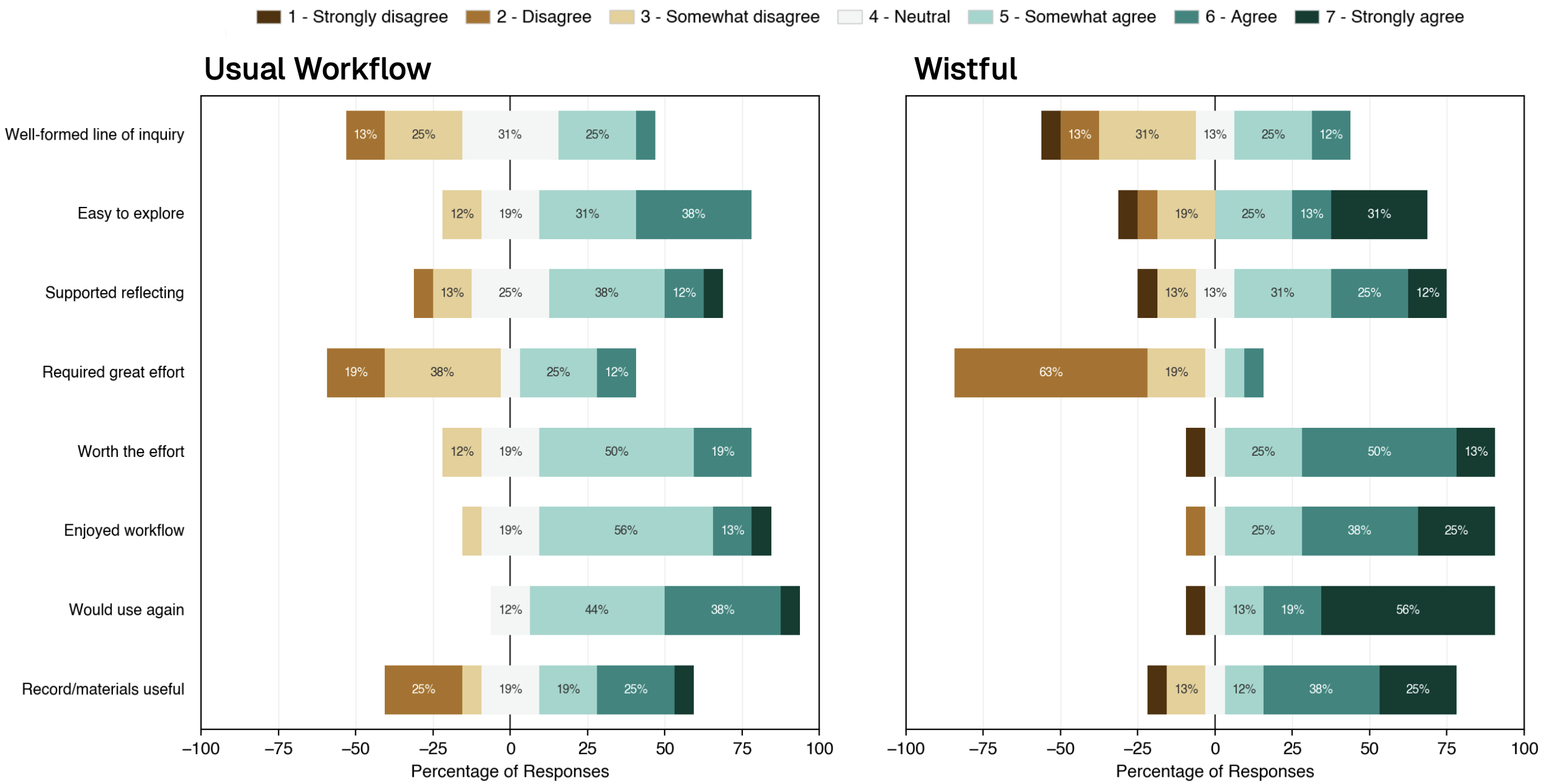}
    \caption{Paired post-task Likert responses for usual workflow (left) and \textsc{Wistful} (right), $N=16$.}
    \Description{Two diverging stacked Likert charts compare participant ratings of their Usual Workflow and Wistful across eight statements on a 7-point scale from strongly disagree to strongly agree. Compared with their usual workflow, participants generally rated Wistful more positively for supporting reflection, being worth the effort, enjoyment, willingness to use again, and usefulness of the resulting record/materials. Participants also largely disagreed that Wistful required great effort. Ratings for forming a well-formed line of inquiry were more mixed.}
    \label{fig:likert_diverging_stacked}
\end{figure}

Participants' post-task ratings were broadly comparable across the two workflows (Fig.~\ref{fig:likert_diverging_stacked}). Descriptively, responses favored \textsc{Wistful} for enjoyment, willingness to use the workflow again, usefulness of the resulting record, and lower effort, while ratings for ease of exploration and forming a line of inquiry were more similar across conditions. This within-participant design grounded participants' reflections in direct comparison with their own established practices, helping them articulate how \textsc{Wistful} changed their experience of scholarly exploration. We use this comparison to contextualize the qualitative and behavioral findings that follow.



\subsection{Researchers Read Through Personal Reference Frames}

\subsubsection{Search Reflected Existing Research Practices}
\begin{figure*}
\centering\includegraphics[width=0.99\linewidth]{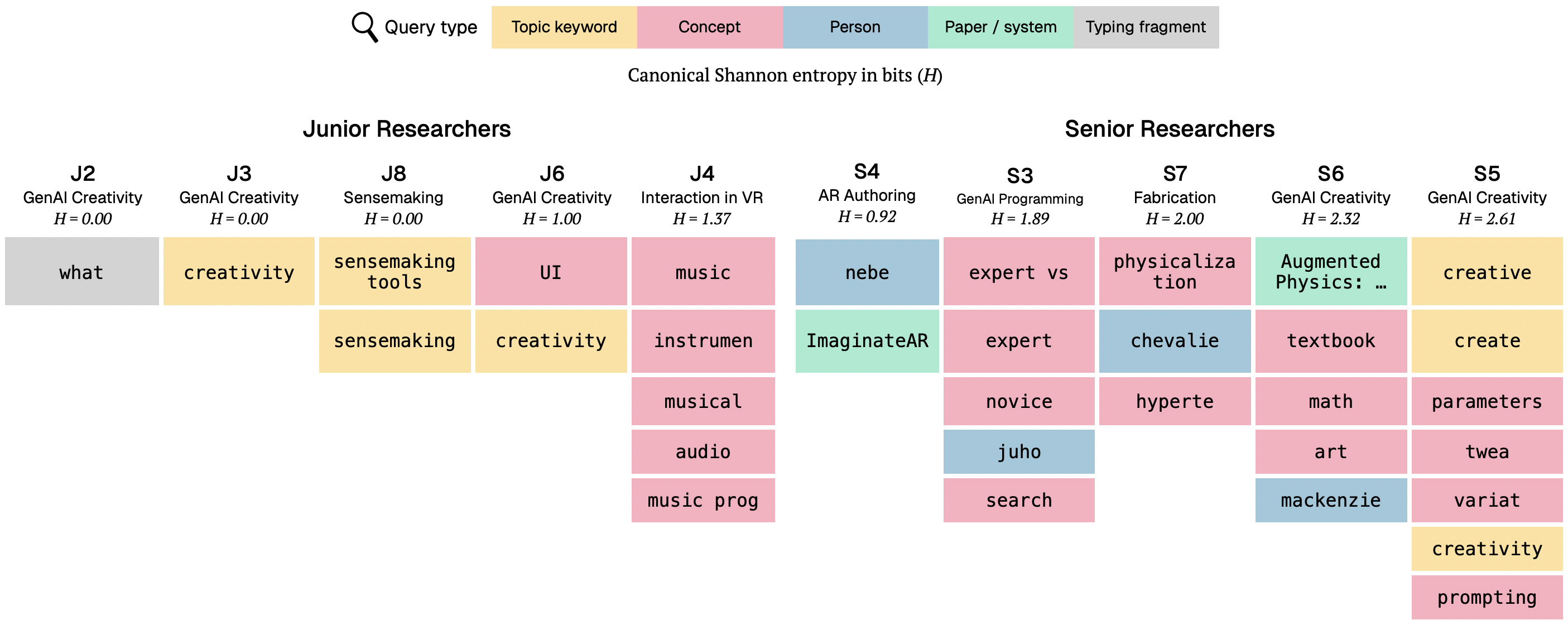}
    \caption{Search queries by participant seniority and query type, ordered by Shannon entropy of distinct queries. Assigned topics are abbreviated. Each tile represents a distinct query; repeated searches are collapsed, and participants who did not search are omitted. Senior searchers had higher median query entropy than junior searchers and more often searched for specific people, papers, or concepts, while juniors more often searched for their assigned topic verbatim.} 
    \Description{Comparison of search queries from five junior and five senior researchers, grouped by query type and labeled with canonical Shannon entropy \(H\). Query types include topic keywords, concepts, people, papers or systems, and typing fragments. Junior researchers generally show lower query entropy, from \(H=0.00\) to \(1.37\), while senior researchers range from \(H=0.92\) to \(2.61\) and use a broader mixture of query types.}
    \label{fig:uistful_search_handles}
\end{figure*}

Participants brought prior experience into how they approached the shared scholarly map. Search made this especially visible (Fig.~\ref{fig:uistful_search_handles}). 10 of 16 participants searched, evenly split between junior and senior researchers. Junior researchers generally searched with the assigned topic, while senior researchers more often searched for specific concepts, people, papers, or systems. After deduplicating repeated queries and collapsing keystroke prefixes, median query entropy was $H=0.59$ bits for junior researchers who searched and $H=2.20$ for senior researchers. Four of five senior searchers searched for an author by name, compared with none of the junior searchers; three seniors also searched for remembered papers or systems, compared with no juniors. S4 explained, \textit{``I know authors in the space, so... I might search for him,''} while S3 searched for a colleague whose recent talk they knew, reasoning, \textit{``Let's get some ideas from what he's been doing.''} J4 was the most diverse junior searcher ($H=1.92$), bringing orchestra recording experience into their VR exploration.

Search was intentionally limited to matching paper titles and author names to encourage exploration through the map and scholarly links. Many participants use AI-assisted search in their usual workflows: 11/16 chose to use an LLM or research-AI assistant during the baseline task (6/8 junior; 5/8 senior). J6 explicitly delegated to AI in this study because the task was \textit{``time-sensitive,''} while S3 used Asta to investigate questions they had already formulated, emphasizing, \textit{``I generate the research topics. I do searches to... find literature.''} The title restriction especially clashed with high-diversity searchers S5 and S3, who had the highest and fourth-highest query entropy, respectively. S3 objected that \textsc{Wistful} was \textit{``not letting me search subjects,''} while S5, who typically uses LLM and regex search over conference abstracts, called title search \textit{``fairly useless''} once they wanted to search for a broader topic keyphrase (``parameter exploration'') across abstracts or full text. For both, the limitation became most apparent once they had a direction they could already name.

\subsubsection{Familiar Landmarks Provided Footholds}
Familiar landmarks provided footholds into the map for 11/16 participants (4/8 junior; 7/8 senior). S6 recognized several colleagues' papers before deciding, \textit{``I'll start with the stuff I'm more familiar with, just to get a sense of what's here.''} Junior researchers more often oriented from papers they had already read or authored. J6 scanned until they found a known paper---\textit{``I want to start exploring from a point that I know''}---while J1 reopened one because \textit{``I attended the talk.''}

\subsubsection{Participants Tested the Map Against Their Own Understandings}
Participants also tested the map against their own understandings of the field. J1 questioned why a paper that \textit{``has nothing to do with AR''} was \textit{``under AR,''} while S5 reacted to the Creativity continent as \textit{``half of the world, somehow.''} S6 similarly asked \textit{``who's dictating those clusters,''} comparing the map against ACM session categories and papers they considered part of creativity. Overall, 12/16 participants questioned or probed the map's organization, including all eight senior researchers and four of eight junior researchers. The shared map gave participants a concrete organization against which to articulate their own understandings of the literature.

\subsection{Familiarity and Personal Interests Influenced the Shape of Scholarly Dérive}
Participants began from an assigned topic on a shared map, but often worked with, around, and sometimes against both the assignment and the map's organization as they pursued their own scholarly dérive. Because \textsc{Wistful} rendered these paths as traces during browsing and preserved them for later analysis, this topical drift became visible both to participants and to us (Appendix Fig.~\ref{fig:all_equirect}). For 5/8 senior researchers, the most-visited continent outside the assignment matched their highest-familiarity alternative; among junior researchers, some departures were instead better explained by personal interests that familiarity ratings did not capture.

\subsubsection{Familiar Ground Acted as a Deliberate Starting Point}
\begin{figure}
\centering\includegraphics[width=0.99\linewidth]{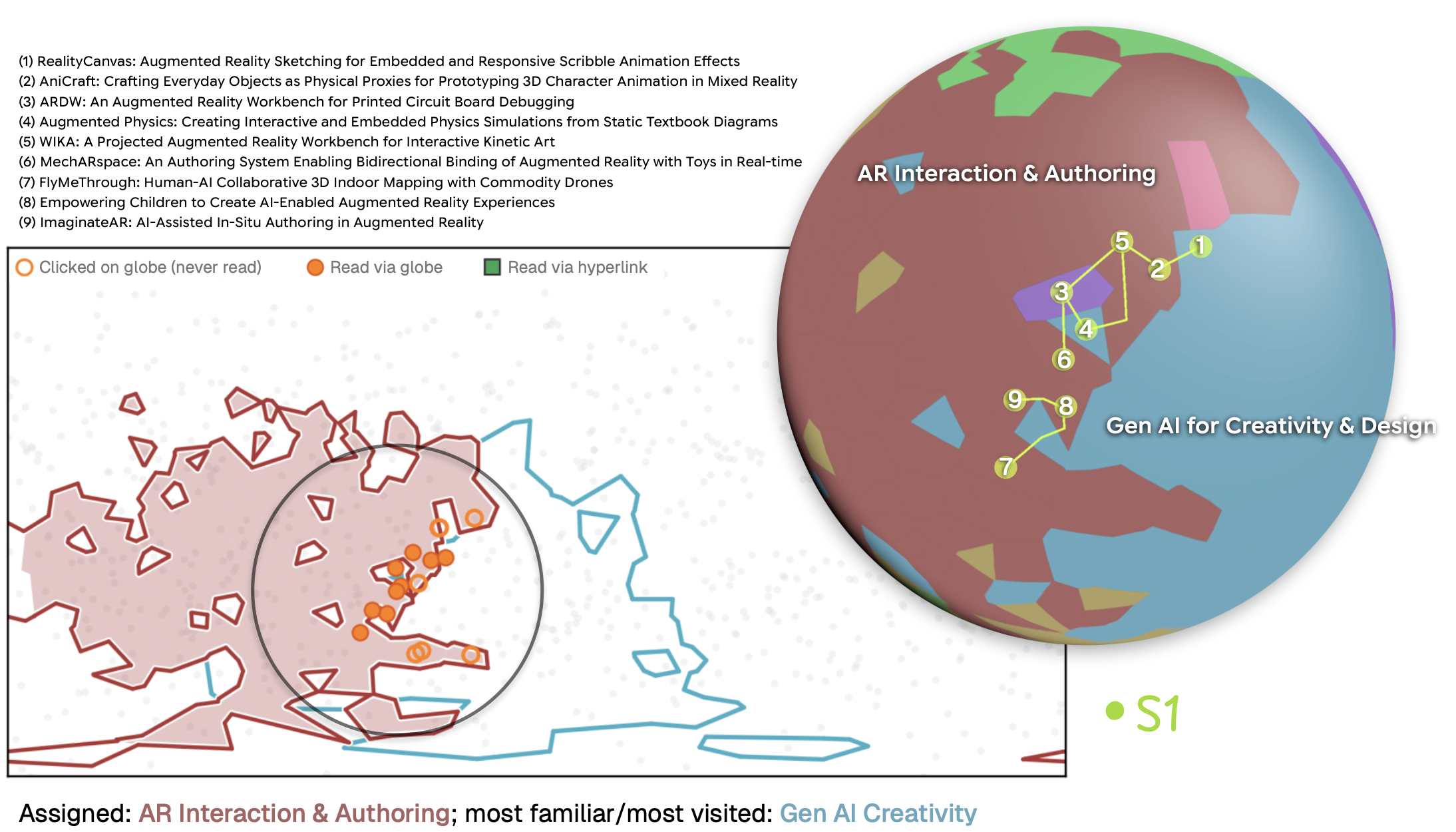}
    \caption{S1's spatial trace. Assigned their least familiar topic, AR Interaction \& Authoring, while GenAI for Creativity \& Design was their most familiar, S1 deliberately began at the boundary between the two and read from familiar work into the assigned continent. Left: the trace projected onto the equirectangular map. Right: trace during globe navigation.}
    \Description{Reading trace for participant S1, who was assigned AR Interaction \& Authoring but spent much of the session in the neighboring Gen AI for Creativity \& Design region. A map at left shows papers S1 clicked or read, while a globe at right shows the numbered sequence of nine visited papers. The trace begins in AR Interaction \& Authoring, moves into Gen AI Creativity, and includes papers reached both from the globe and through hyperlinks.}
    \label{fig:s1_globe}
\end{figure}

Familiarity influenced where some participants read within an assigned topic. For the three senior researchers assigned the less-familiar AR Interaction \& Authoring topic while GenAI for Creativity \& Design was highly familiar, exploration clustered along and sometimes crossed the GenAI-facing boundary. Among their assigned AR reads, 6/6 for S1, 7/9 for S4 (Fig.~\ref{fig:j7_s4_hyperlinking}, right), and 4/6 for S8 fell within the quarter of AR papers closest to GenAI Creativity; all three also opened papers in that continent. S1 made this strategy explicit (Fig.~\ref{fig:s1_globe}), explaining, \textit{``my core area is this blue area, like GenAI, so I want to see what are on the boundaries here.''} This echoed their usual tendency to favor papers \textit{``in the boundary of the different conference venues.''} Afterward, they described the session as \textit{``exploring the unknown territory, but starting from the territory that I know.''}

J2 showed a similar boundary strategy despite already being assigned their most familiar topic. They looked \textit{``along this border''} for papers with \textit{``a bit of both,''} favoring the programming-facing edge because \textit{``I don't really do fabrication stuff.''} They sought work where someone \textit{``coming from this other spot''} had published within their assigned area, later observing that \textit{``basically every paper here is something that I would either be interested in or have already read.''}

\subsubsection{Personal Interests Pulled Reading Across Research Boundaries}
\begin{figure}
\centering\includegraphics[width=0.99\linewidth]{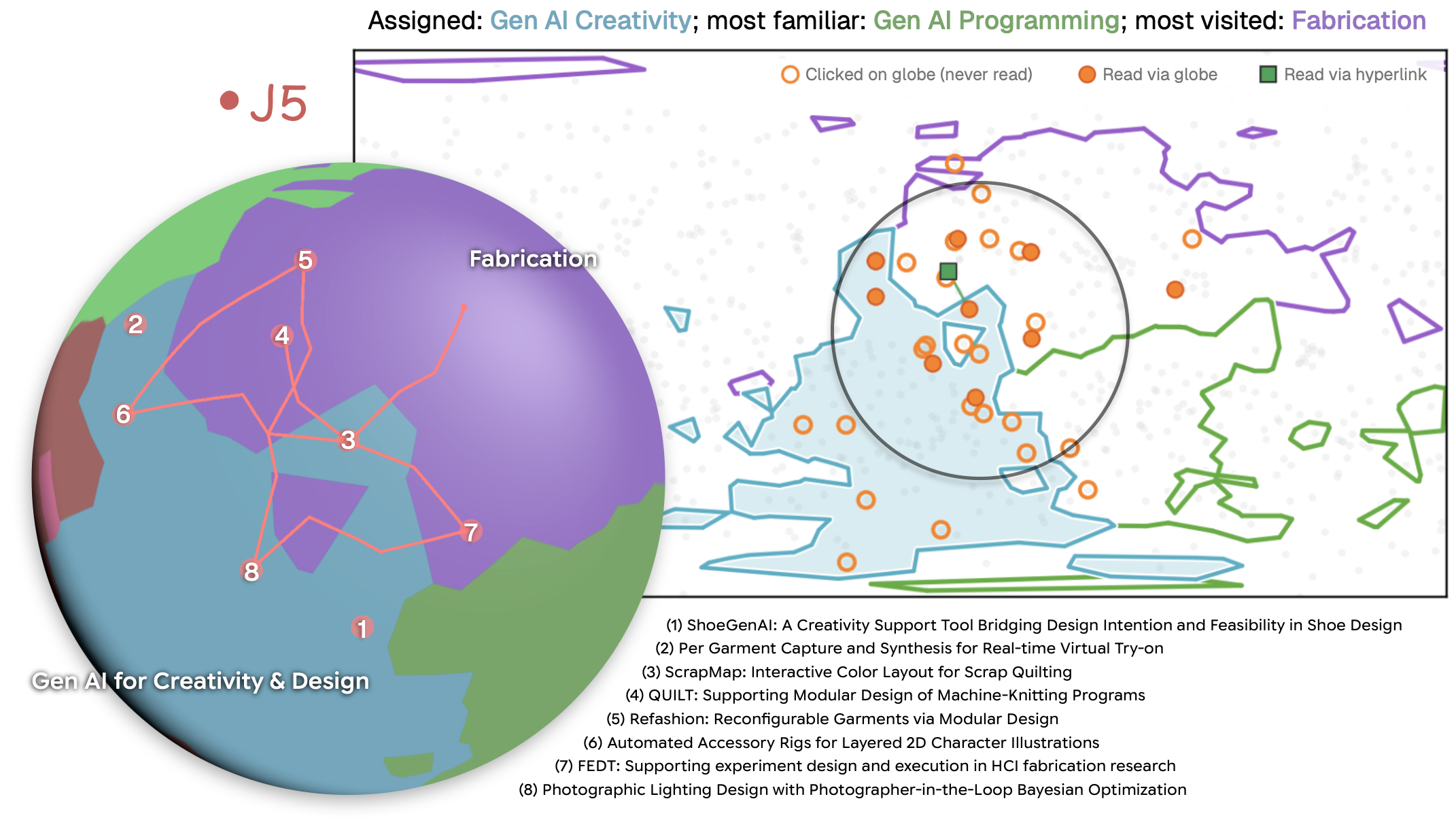}
    \caption{J5's spatial trace. Although assigned GenAI for Creativity \& Design, much of their exploration extended into Fabrication.}
    \Description{Reading trace for participant J5, who was assigned Gen AI Creativity, was most familiar with Gen AI Programming, and visited Fabrication most. A map shows clicked and read papers clustered around Gen AI Creativity and Fabrication, while a globe view shows an eight-paper path moving repeatedly between those two regions.}
    \label{fig:j5_globe}
\end{figure}

Personal interests also directed paths in ways that research familiarity alone did not explain. J5 had been actively thinking about GenAI for Programming before the study, asking during their baseline task, \textit{``What if I was thinking about this research topic, like, 3 days ago and I have notes on it?''} In \textsc{Wistful}, however, they primarily read garment, quilting, and refashioning work across GenAI Creativity and Fabrication (Fig.~\ref{fig:j5_globe}). J5 described intentionally keeping sewing and fashion design separate from their research, preferring connections between different parts of their life to arise unexpectedly rather than deliberately seeking them out. Yet this personal interest surfaced in their reading trace without being an explicit research goal, shaping what they noticed and followed.

S7 showed a similar crossover through crochet. Assigned their least familiar topic, Fabrication, while GenAI Creativity was their most familiar, they looked for work \textit{``at the intersection of two of these things''} and opened \textsc{StitchFlow}. The Delaunay edges shown on the map revealed that its citation neighborhood led into Fabrication (Appendix Fig.~\ref{fig:stitchflow_map}). After seeing those links, S7 reacted, \textit{``That looks great! ... Let's jump in here,''} adding that crochet was \textit{``one of [their] hobbies''} and \textit{``super computational.''} They then followed a citation into their assigned Fabrication area. Their path shows how familiar research areas and personal interests could together draw participants into less familiar territory.




\subsection{Readers Composed Paths Across Semantic and Citation Structures}


The map and hyperlinks exposed different relationships within the same corpus. The map organized papers through continuous semantic proximity, while hyperlinks exposed citation relationships that could span distant parts of that terrain. Participants recognized the map as one possible organization of the literature. S6 observed that \textit{``a lot of it comes down to who's dictating those clusters, and how much faith do I have in them,''} noting that knitting papers they considered creativity \textit{``might have been missed on the map.''} S4 similarly described the map as \textit{``a different view''} of relationships they would usually approach through citations.

\begin{figure}
\centering\includegraphics[width=0.99\linewidth]{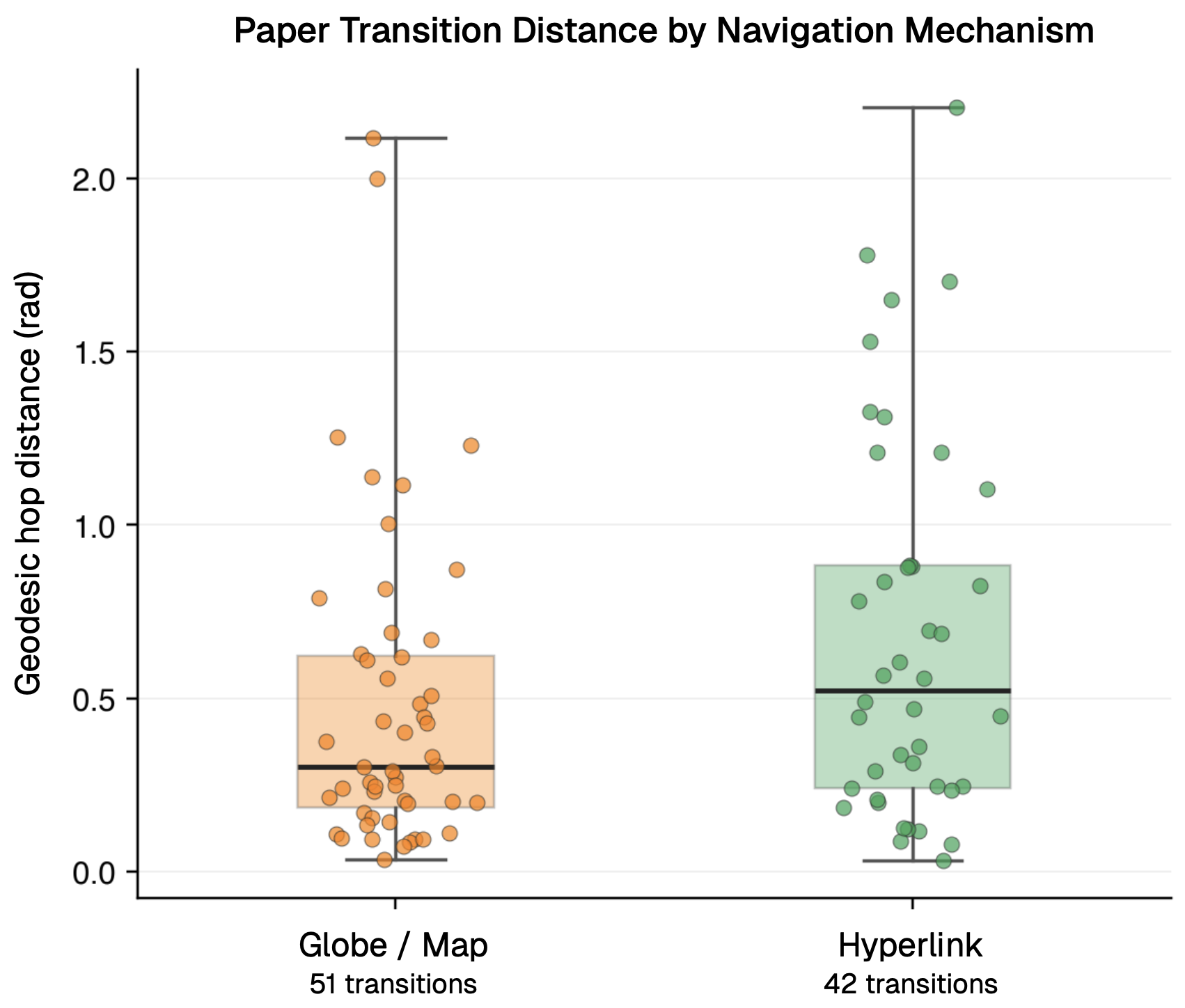}
    \caption{Transition distance between consecutive unique paper first-opens from the map (51 transitions) or a hyperlink (42 transitions). Search transitions are omitted and distance is geodesic on the globe. Hyperlinks generally carried readers farther across the semantic landscape than map navigation.}
    \Description{Box-and-jitter plot comparing geodesic paper-transition distance for two navigation mechanisms. Globe/Map includes 51 transitions with a median distance of about 0.30 radians, while Hyperlink includes 42 transitions with a higher median of about 0.52 radians. Hyperlink transitions are generally more dispersed and tend to span longer distances, although both mechanisms include a few transitions above 2 radians.}
    \label{fig:transition_distance}
\end{figure}

Hyperlinks gave readers another structure through which to move alongside the semantic map. Inspired by expressive range analysis~\cite{expressiveRange, exprRangeCreative}, we examine how far these navigation mechanisms carried readers across the same map. Among first-opens of new papers, excluding search transitions, map transitions had a median geodesic distance of $0.30$ radians, compared with $0.52$ for hyperlinks (Fig.~\ref{fig:transition_distance}). Hyperlinks also crossed continent boundaries more often (45\% vs. 27\%). Among the 11 participants who used both, eight had a greater mean hyperlink distance. Hyperlinks therefore more often carried readers beyond the local semantic neighborhoods they traversed on the map. Navigation mix was similar across career stages: hyperlinks accounted for a median 33\% of junior researchers' reads and 37\% of senior researchers' reads.

\begin{figure}
\centering\includegraphics[width=0.99\linewidth]{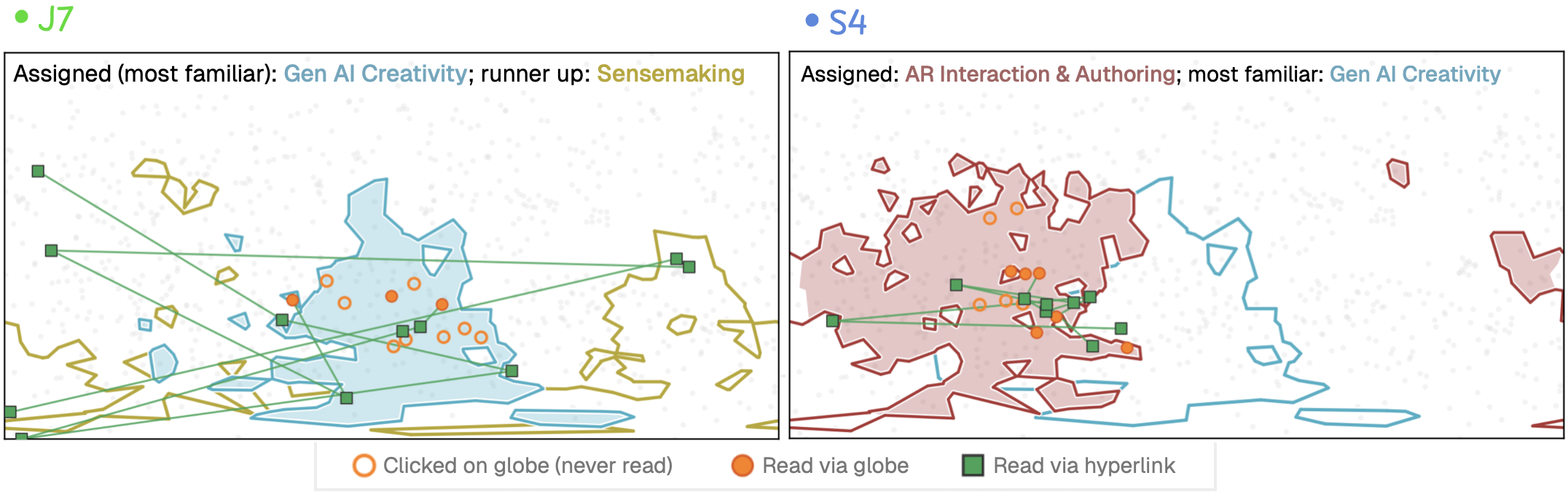}
    \caption{Spatial traces of J7 and S4 projected onto the semantic map. J7 and S4 together accounted for 20/42 hyperlink first-opens, which often connected more distant regions of the map.}
    \Description{Side-by-side reading traces for junior participant J7 and senior participant S4. J7, assigned to a topic where they were most familiar with Gen AI Creativity and second-most familiar with Sensemaking, follows several long hyperlink jumps across the map. S4, assigned AR Interaction \& Authoring and most familiar with Gen AI Creativity, stays more concentrated within the assigned AR region. Orange markers indicate globe interactions; green squares and lines indicate papers reached through hyperlinks.}
    \label{fig:j7_s4_hyperlinking}
\end{figure}

J7 and S4's paths make this pattern especially visible (Fig.~\ref{fig:j7_s4_hyperlinking}). Together, they accounted for 20 of the 42 hyperlink first-opens, and both made roughly 80\% of their paper opens through hyperlinks. Yet J7's reads spanned five continents ($H=1.99$), while S4's remained concentrated across three ($H=0.87$). J7, who made 11 of their 14 first-opens through hyperlinks, noted that the two papers they ended up \textit{``kind of merging''} for their line of inquiry were \textit{``in separate parts of the map.''} Their usual workflow began with keyword- or prompt-based search through Asta or Google Scholar, followed by references from an initial set of results. Reflecting on their baseline task, J7 described how they could \textit{``get fixated on one specific direction.''} In \textsc{Wistful}, they instead encountered related papers \textit{``not in that initial search,''} ultimately describing them as part of \textit{``the same line of story, but ... just a different angle of it.''}

Readers also shifted between structures as an inquiry changed course. Twelve of 16 participants opened papers through both the map and hyperlinks, and eight returned to the map after at least one hyperlink open. After following references, J7 said they had \textit{``hit, like, a dead end''} and had \textit{``gone back to the map.''} J1 similarly described getting \textit{``sidelined''} before deciding to \textit{``go back ... to the top level.''} Reflecting afterward, they described knowing when a rabbit hole had gone too far and when to \textit{``go back another one.''} Readers could therefore follow connections away from a local semantic neighborhood while retaining the map as a place to reorient and begin another path.



\subsection{Reading Traces Supported Orientation, Reflection, and Exchange}

Reading traces are partial records of an exploration, shaped by what readers noticed and followed. They do not capture a scholarly sensibility in full, but can preserve some of its expression in a given session. Participants used traces to keep track of where they had been, preserve context around what they encountered, and compare their paths with those of others, while pointing toward opportunities for deeper reflection.

\subsubsection{Traces Supported Orienteering During Exploration}
Participants could revisit their accumulating traces while still reading. J1 opened \textit{My Trace} to \textit{``take a look at what I've been doing.''} They recalled, \textit{``I started with these two things, and then I was like, no! Too out of scope, let's back out,''} before observing that they were now \textit{``more in AR land.''} Later, J1 described how they had \textit{``made an M''} and could \textit{``linearly see where I went.''} J3 similarly returned to \textit{My Trace} to check which papers they had already viewed before returning to search. The trace helped readers keep track of where they had been and, for J1, reorient their exploration.

\subsubsection{Field Notes Preserved Context Around Reading}
In the usual-workflow condition, 14/16 participants used the provided Google Doc, often to collect titles, links, and brief labels; 10/16 wrote at least one note in their own words. In \textsc{Wistful}, 14/16 used the prompted Field Notes, and 13/16 wrote at least one Field Note in their own words. Across coded segments, 61\% of Field Note lines were in participants' own words, compared with 25\% of baseline document segments; 71\% of baseline segments were instead labels, links, or references. 

J1 contrasted this with their usual workflow, where there was \textit{``no scaffolding for me to actually do reflection,''} describing \textsc{Wistful} as making \textit{``forming connections the first-class citizen of this.''} They also found the persistent Field Note affordance \textit{``very compelling,''} because \textit{``the sticky note is already there, you must use it.''} J3 explained that they usually \textit{``keep track of the title''} in their notes, but because \textsc{Wistful} \textit{``already tracks all the papers I've looked at,''} they no longer needed to save each paper as a link. J6 similarly described saving every opened paper to Zotero yet still being able to \textit{``get lost''} when Zotero failed or many papers accumulated in browser history. 

\subsubsection{Revisiting Traces Pointed Toward a Longer Time Horizon}
Reflection Notes saw lighter use during the study: seven participants wrote 11 notes, including links between papers, reminders, follow-up questions, and emerging ideas. J2, who used neither Field nor Reflection Notes, explained that they could simply \textit{``remember everything that happened''} during a 15-minute session, but imagined the notes becoming useful after \textit{``I went to sleep and I came back to it.''} In a fuller reading session, they imagined writing another note after finishing a paper using \textit{``the reflection thingy.''} J1 similarly distinguished the more \textit{``ephemeral''} Field Notes from Reflection Notes that \textit{``persist,''} imagining that \textit{``at the end of my lit search, I would do a reflection on older reflections.''}

J8 did not use Reflection Notes, but described revisiting a trace as \textit{``kind of like a reflection... on your past search.''} While discussing whether something aligned with their \textit{``research taste or... research interest,''} they suggested that the trace could help \textit{``visualize that intuition.''} J2 likewise thought a trace could help them avoid \textit{``retrac[ing] my steps,''} while J8 planned to continue more in-depth research \textit{``based on the trace I had.''} These accounts point toward reflection becoming more useful as traces accumulate beyond a single short session.

\subsubsection{Shared Traces Supported Comparison and Exchange}

Viewing shared traces made differences between reading paths visible (Fig.~\ref{fig:workflow_2}). Junior researchers' opened-paper traces were somewhat more dispersed across continents than seniors' ($H=1.16$ vs.\ $0.62$). J7 noticed, \textit{``It's interesting that some of the senior ones are, like, shorter traces. Like, they know exactly what they want to do. Oh my god, I'm, like, all over the... this is me?''} J6 similarly contrasted their trace---\textit{``everywhere''}---with senior traces that appeared more \textit{``aggregated,''} while also emphasizing \textit{``a lot of individual variants''} in exploration.

Shared traces could also lead readers toward work they had not encountered themselves. While reviewing S4's trace, S8 commented, \textit{``almost miss these 3 papers!! thanks for your trace!! super helpful!''} Because viewing and commenting were prompted by the study protocol, we treat this as an example of how another reader's path could become a useful starting point rather than evidence of spontaneous exchange.

Participants differed in what they considered worth passing on. S1 imagined a student \textit{``starting off from my own trace''} and comparing how their exploration developed. Others valued the papers more than the path itself. S2 preferred to \textit{``export a list of paper within the trajectory''} because its \textit{``ordering... doesn't really mean something''} specific, while S8 similarly imagined sending a list of papers rather than the trace.

Shared traces also exposed limits to what a trace alone made legible. We did not ask participants to curate their published traces into explanations for another reader, but S6 read S4's trace through a mentorship lens. They noted that there were \textit{``not a lot of notes here to go off of''} and questioned how useful the graph would be to an undergraduate \textit{``without any annotation.''} S4 similarly reflected that although the trace had \textit{``recorded everything I was doing,''} their process was \textit{``not very linear,''} and neither the trace nor their usual Google Doc would have \textit{``told the student much about my process.''} These cases suggest that when a reading path becomes something others may inherit, the choices behind it may require additional context.

\section{Discussion}
Our findings show that researchers approached \textsc{Wistful}'s open scholarly landscape through plural, idiosyncratic orientations shaped by accumulated experience, which we describe as \textbf{\textit{scholarly sensibilities}}. These differences appeared in the paths participants enacted through reading, as well as in what they chose to record explicitly in their notes. Preserving these choices and reflections as \textbf{\textit{creative reading traces}} made otherwise ephemeral expressions of scholarly sensibility available for analysis, while participants themselves primarily used their own and others' traces for orientation, recollection, comparison, and exchange. We now consider what these findings suggest for supporting the expression and development of scholarly sensibilities as AI-mediated information ecosystems increasingly shape how researchers read and, crucially, how those sensibilities form.

\subsection{From Sensemaking to Sensibility-Making}

We extend accounts of professional vision and scientific taste by drawing attention to the broader sensibilities through which researchers find their way through literature. Professional vision describes socially learned disciplinary ways of seeing~\cite{scientistsProfessionalVision}, while scientific taste concerns which questions and directions seem worth pursuing~\cite{aQuestionOfTaste}. Scholarly sensibilities encompass these forms of expertise alongside the wider experiences and interests that draw a researcher's attention. Gamboa and Andersen describe sensibility as a ``skin'' formed through experience, closer to a way of \textit{being} than a bounded way of \textit{working}~\cite{Sensibilities}. Bush writes that our steps in ``creating or absorbing material of the record'' proceed through the \textit{senses}~\cite{AsWeMayThink}. Over time, experience settles into \textit{sensibilities} that shape how researchers negotiate that record.

\textsc{Wistful}'s creative reading traces gave expressions of these sensibilities a visible form. Senior researchers described gaining a \textit{``picture of the landscape''} over their careers, and this accumulated familiarity was visible in several traces, with participants concentrating their reading in the part of an unfamiliar continent nearest their most familiar continent. That ``skin'' also carried experience beyond formal research: J5's interest in sewing and fashion shaped what they followed despite keeping it separate from their research, while J1 described ideas developing through informal conversations and everyday encounters before a literature search began. Sensibilities thus travel with the researcher across research and everyday life, shaping their work both explicitly and implicitly.

Sensibility can guide reading before it can be fully articulated. J2 described being \textit{``not able to articulate [their interests] very well to other people, but... can kind of tell it myself,''} sensing \textit{``there is a thing that's there''} without yet having the \textit{``sentence or a short paragraph''} to express it. Information-seeking research similarly recognizes that needs can precede their articulation as queries~\cite{questionNegotation}, while expertise shapes the language available for searching~\cite{queryFormulation}. This appeared in the way participants searched: senior searchers, despite being assigned their least familiar topic, more often drew on names, papers, and concepts, while junior searchers, despite being assigned their most familiar topic, more often used the assigned topic itself. Subramonyam et al.'s ``gulf of envisioning'' describes a related challenge in prompting, where emerging intentions must be translated into language an AI system can act on~\cite{gulfEnvisioning}. For scholarly sensibility, this gulf may arise before a researcher can name a direction that already feels worth following. S3, our most senior participant, used an LLM in their usual workflow, searched extensively in \textsc{Wistful}, and criticized its limited search, illustrating the more articulated end of this process: \textit{``I generate the research topics. I do searches to... find literature.''}

This suggests an opportunity for \textit{sensibility-making}, building on prior work that frames creative reading as a process through which readers develop the orientations they bring to future work~\cite{CreativeReading}. As AI systems increasingly infer researchers' interests, reading augmentation systems might also help researchers recognize and give voice to their own. \textsc{Wistful} let participants follow what drew them before fully articulating it and preserved those choices in traces, creating material for later reflection.


\subsection{Scholarly Dérive Under Pressure}

Inspired by Debord's \textit{dérive}, meaning ``drifting,'' we use scholarly dérive to describe playful movement through literature shaped by both the attractions of the terrain and the reader's own sensibilities~\cite{Debord}. Gamboa and Andersen similarly draw on dérive as a practice of wandering while remaining attentive, where a ``tolerance for drifting'' becomes part of an evolving sensibility~\cite{Sensibilities}. Scholarly dérive gives sensibility room to act before a direction is fully articulated. We intentionally designed both \textsc{Wistful} and the study task to leave room for this kind of drift. Even within our 15-minute tasks, participants followed personal interests and familiar terrain into unplanned directions that became visible in their traces.

Dérive also requires time for directions to be followed before their value is clear. Open-ended discovery can depend on stepping stones whose importance only becomes apparent later~\cite{whyGreatness}, while scientific creativity depends on conditions that leave room for uncertainty and longer-horizon inquiry~\cite{scientificCreativityInstitutional}. Participants described this tension directly. S3 attributed their ability to connect \textit{``a thousand things''} to having thought about them \textit{``for many years,''} while S6 argued that spending \textit{``a whole day, or even a whole week, or even a whole month''} reading should count as academic labor. Yet S1 described deadlines determining whether an idea becomes \textit{``crazier or not,''} and S8 felt increasingly \textit{``less likely to explore totally new ideas''} as they became more senior. Participants worried that AI could further accelerate this process. S4 associated skipping reading with \textit{``not really developing [students'] own taste,''} while J5 described research choices becoming \textit{``watered down''} when \textit{``someone else made [the decision]... or, like, an agent.''} For junior researchers especially, compressing these experiences may also compress opportunities to discover what draws them and develop their own sensibilities.

Our short study also points toward how traces might connect these different timescales. Although each task captured only 15 minutes, traces still bore the imprint of much longer histories of reading and experience. Participants then imagined those traces extending forward: J2 anticipated the trace becoming useful after \textit{``I went to sleep and... came back to it,''} J1 imagined reflecting on \textit{``older reflections,''} and J8 described continuing deeper research \textit{``based on the trace''} they had made. Creative reading traces can thus preserve a moment shaped by accumulated sensibility while providing a scaffold for inquiry that continues beyond it. Systems could support scholarly dérive across time by preserving paths for later return.



\subsection{Making the Scholarly Map Come to Life}
\begin{figure*}
    \centering\includegraphics[width=0.99\linewidth]{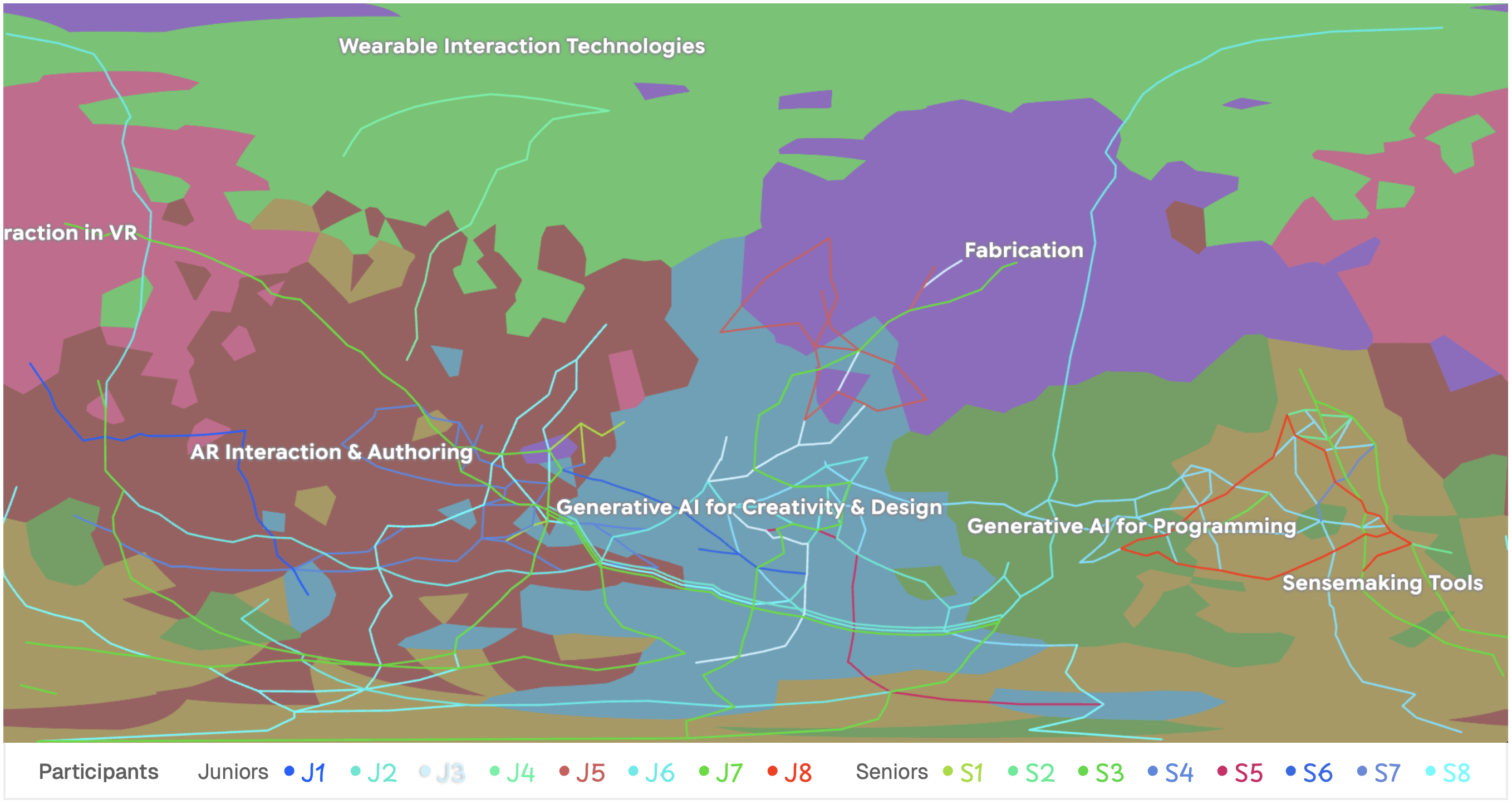}
    \caption{All participant traces projected onto the equirectangular map across seven continents, showing a plurality of readings; trace colors were selected by participants during the study.}
    \Description{Map of the UIST scholarly landscape with seven labeled research regions and reading traces from 16 participants. Junior participants J1–J8 and senior participants S1–S8 are shown as differently colored paths. Traces vary from localized exploration within one region to long paths crossing several regions, with substantial activity around AR Interaction \& Authoring and Generative AI for Creativity \& Design.}
    \label{fig:all_participant_traces}
\end{figure*}
\textsc{Wistful} builds on \textsc{Artographer}'s use of an intentionally constructed spatial map as both a terrain for exploration and an object of interpretation, where participants interrogated the computational organization of art space against their own understanding of the domain~\cite{Artographer}. We observed the same dynamic in scholarly reading: 12/16 participants questioned or probed the map's organization, including all eight senior researchers. Senior researchers may have had more established reference points against which the map could appear surprising or wrong, while junior researchers could use the same representation to begin forming an overview of the field. Senior participants explicitly imagined this pedagogical role: S7 described the bounded map as \textit{``a really productive constraint''} that could keep junior researchers from being sent \textit{``off into the void,''} while S5 imagined using its continents to introduce students to the breadth of UIST. For emerging researchers, the map can help them learn the shape of a field while also shaping the field they learn to see.

Bey imagines nomads charting their course by \textit{``strange stars,''} including \textit{``luminous clusters of data in cyberspace,''} layered beneath a \textit{``1:1 map of the creative imagination, aesthetics, values''} until the \textit{``resultant grid comes to life''}~\cite{TAZ}. \textsc{Wistful}'s UMAP landscape is likewise one computational representation of the literature, constructed from patterns in the data rather than a definitive map of the field. As readers moved through it, they layered their interests, judgments, and associations onto that representation, bringing the landscape to life through creative reading~\cite{CreativeReading}. Citation links offered another map of the literature, inscribed by paper authors through the scholarly relationships they chose to anoint, and allowed readers to ``teleport'' across the semantic terrain when those relationships cut across spatial proximity. Their traces preserve yet another layer by recording each reader's path through these overlapping structures, effectively \textit{writing onto}---and at times \textit{subverting}---the shared map without changing its underlying structure, and leaving that interpretation behind for others to see (Fig.~\ref{fig:all_participant_traces}).

\textsc{Wistful} keeps its computational organization and the corpus it organizes available for readers to traverse rather than choosing a path for them. Its Delaunay edges and citation links offer possible routes, building on hypertext traditions that treat navigation as authorship~\cite{deathOfTheAuthor, becomingHypertext}. This principle extends beyond literal maps: embeddings, rankings, recommendations, and LLM-mediated search likewise organize what becomes visible and adjacent. Such systems should leave room for readers to navigate and contest these organizations rather than silently accepting the paths this mediation privileges~\cite{hypertextualFriction}.




\subsection{Creative Reading Traces as Material for Reflection and Exchange}

Scholarly sensibilities are also developed and expressed socially. J1 described their research process as \textit{``super social''} and papers as \textit{``conversation points,''} later arguing that the study was \textit{``underselling''} \textsc{Wistful}'s social component. J2 could already recognize papers that interested them, but wanted to articulate that orientation well enough that another person could say, \textit{``I know whether or not [J2] would be interested in this.''} Part of developing a scholarly sensibility may be learning to make it legible to others, so that colleagues can bring new papers or people into one's orbit. Shared traces offer another way for these orientations to become socially available before they can be fully put into words.

We saw this begin when participants read one another's traces. J2 was especially interested in S1's path knowing they were a senior researcher and considered adopting aspects of their strategy (Fig.~\ref{fig:teaser}), while S8 found papers through S4's trace. Traces let researchers read not only the literature, but one another's \textit{readings} of it, making differences in attention and approach available for comparison and learning~\cite{cogAppren}. Bush imagined scholars deliberately leaving trails so that disciples might inherit ``the entire scaffolding by which [knowledge] was erected''~\cite{AsWeMayThink}. Our participants were not asked to construct such trails for others, yet even brief traces of ordinary reading carried parts of that scaffolding. What remained harder to inherit was the rationale behind the path, which points to opportunities for traces to develop alongside conversation and explanation.









\subsection{Limitations \& Future Work}
Our study captures only brief expressions of scholarly sensibilities within a fixed UIST corpus and 15-minute tasks. \textsc{Wistful} was designed to provide a shared canvas on which existing sensibilities could become visible through reading traces, rather than to produce or change those sensibilities itself. We therefore cannot make causal claims about career stage or sensibility development, and traces remain partial records that preserve movement more readily than the judgments behind it.

At the same time, our findings point toward designing for \textit{sensibility-making}. Participants imagined returning to traces over longer periods, using them to reflect on prior reading, compare with others, and continue inquiry. Longitudinal deployments could study how traces accumulate across months or years, whether recurring patterns help researchers recognize and articulate what draws them, and how peers and mentors interpret and discuss those traces. Future work could also compare how different forms of AI-mediated reading shape these trajectories, and whether traces can help readers remain conscious of the interests and judgments guiding them. Such systems could support sensibility-making by making sensibilities visible as they are expressed through reading, helping readers recognize their own orientations and reflect on how they develop over time.

\section{Conclusion}
\begin{quote}
    \textit{The gestalt, Dear Brutus, is not in the stars but in our selves.}
\end{quote}
\begin{flushright}---Ted Nelson, \textit{Dream Machines}~\cite{dreamMachines}\end{flushright}

Researchers bring more to reading than a need for information. Through \textsc{Wistful}, we showed how scholarly sensibilities become visible in the paths researchers make through literature, and how creative reading traces can preserve these fleeting expressions for reflection and exchange. As scholarly systems increasingly model readers and personalize what they encounter, we argue for designing beyond sensemaking toward \textit{sensibility-making}, supporting readers in recognizing and developing their own ways of seeing. The stars on these maps may be data, but the constellations readers draw among them are their own.


\begin{acks}
We thank Kosa Goucher-Lambert, Hugh Dubberly, and other members of the knowledge graph discussion group, as well as Ronen Tamari and Wesley Finck of the Cosmik Network, for thoughtful discussions and inspiration throughout the development of Wistful. We are especially grateful to Samia Menon, Tim Aveni, and James Smith for their support and contributions since the earliest stages of this project, and to Andy Matuschak for his insights on creative reading. Finally, we thank our study participants for generously sharing their time and perspectives on their research practices.
\end{acks}

\bibliographystyle{ACM-Reference-Format}
\bibliography{refs}


\clearpage
\appendix
\onecolumn

\section{Appendix}

\subsection{Formative Study Participants}
\label{app:formative-participants}

\begin{table}[H]
\centering
\caption{Formative study participant backgrounds, Wikipedia use, rabbitholing frequency, and starting articles.}
\label{tab:formative-participants}
\begin{tabular}{lllll}
\toprule
\textbf{ID} & \textbf{Background} & \textbf{Wikipedia use} & \textbf{Rabbitholing} & \textbf{Starting article} \\
\midrule
P1 & Graduate student & Occasionally & Rarely & X-Men \\
P2 & PhD student & Occasionally & Very often & California quail \\
P3 & PhD student & Occasionally & Rarely & Xiaohongshu \\
P4 & Homemaker & Regularly & Very often & Organized crime \\
P5 & Graduate student / editor & Regularly & Very often & Whole Earth Catalog \\
P6 & Undergraduate student & Occasionally & Sometimes & Monte Carlo method \\
P7 & UX designer & Regularly & Sometimes & Africa (song) \\
P8 & Graduate student & Occasionally & Sometimes & Project Hail Mary \\
P9 & Firefighter / paramedic & Regularly & Very often & Jonestown \\
\bottomrule
\end{tabular}
\end{table}

\FloatBarrier

\subsection{\textsc{Wistful} Study Participants}
\label{app:participants}

\begin{table}[H]
\centering
\footnotesize
\setlength{\tabcolsep}{8pt}

\caption{Participant backgrounds and task assignments. Participant IDs beginning with J denote junior researchers and IDs beginning with S denote senior researchers. Junior researchers were assigned their two most familiar continents and senior researchers their two least familiar continents based on pre-study screener ratings.}
\label{tab:participants}

\begin{tabular}{llllll}
\toprule
\textbf{ID} & \textbf{Career stage} & \textbf{HCI exp.} & \textbf{System order} & \textbf{Familiarity order} & \textbf{Assigned topics (T1, T2)} \\
\midrule
J1 & Teaching faculty & 3--5 y & Usual, Wistful & more, less & GenAI creativity, AR authoring \\
J2 & Junior PhD (1--3) & 3--5 y & Wistful, Usual & more, less & GenAI creativity, Interaction in VR \\
J3 & Master's student & $<$3 y & Usual, Wistful & less, more & Interaction in VR, GenAI creativity \\
J4 & Postbaccalaureate & $<$3 y & Wistful, Usual & less, more & Interaction in VR, Fabrication \\
J5 & Industry researcher & $<$3 y & Usual, Wistful & more, less & GenAI programming, GenAI creativity \\
J6 & Junior PhD (1--3) & $<$3 y & Wistful, Usual & more, less & GenAI creativity, Sensemaking \\
J7 & Master's student & $<$3 y & Usual, Wistful & less, more & Sensemaking, GenAI creativity \\
J8 & Junior PhD (1--3) & $<$3 y & Wistful, Usual & less, more & Sensemaking, GenAI creativity \\
S1 & Industry research scientist & 6--10 y & Usual, Wistful & more, less & Wearable interaction, AR authoring \\
S2 & Senior PhD (4+) & 6--10 y & Wistful, Usual & more, less & Sensemaking, Interaction in VR \\
S3 & Full professor & $>$20 y & Usual, Wistful & less, more & Interaction in VR, GenAI programming \\
S4 & Full professor & $>$20 y & Wistful, Usual & less, more & AR authoring, Wearable interaction \\
S5 & Full professor & $>$20 y & Usual, Wistful & more, less & Fabrication, GenAI creativity \\
S6 & Assistant professor & 3--5 y & Wistful, Usual & more, less & GenAI creativity, AR authoring \\
S7 & Assistant / adjunct professor & 6--10 y & Usual, Wistful & less, more & Wearable interaction, Fabrication \\
S8 & Senior PhD (4+) & 6--10 y & Wistful, Usual & less, more & AR authoring, Sensemaking \\
\bottomrule
\end{tabular}
\end{table}

\FloatBarrier

\subsection{Post-Task Reflection Questionnaire}
\label{app:post-task-questionnaire}

Participants completed a post-task reflection form after each of the two exploration tasks. Junior and senior researchers received separate forms with task wording adapted to their respective study roles. After the first task, participants completed the corresponding task block and paused before continuing the study; after the second task, they completed the second block and the remainder of the form. Unless otherwise noted, closed-response items used a 7-point scale. Familiarity items ranged from 1 (\textit{Not at all familiar}) to 7 (\textit{Extremely familiar}), and agreement items ranged from 1 (\textit{Strongly disagree}) to 7 (\textit{Strongly agree}).

\subsubsection{Shared Header}

\begin{enumerate}
    \item \textbf{Participant ID}
    \item \textbf{Which task did you complete first?}
    \begin{itemize}
        \item Usual Workflow
        \item Wistful
    \end{itemize}
\end{enumerate}

\subsubsection{Junior Researcher Form}

The following block was completed once for each task.

\paragraph{Topic and Familiarity}

\begin{enumerate}
    \item \textbf{Which topic did you explore for this task?}

    \item \textbf{Before this task, how familiar were you with this research topic?} \\
    1 (Not at all familiar) -- 2 -- 3 -- 4 -- 5 -- 6 -- 7 (Extremely familiar)

    \item \textbf{After this task, how familiar were you with this research topic?} \\
    1 (Not at all familiar) -- 2 -- 3 -- 4 -- 5 -- 6 -- 7 (Extremely familiar)
\end{enumerate}

\paragraph{Task Reflection}

\noindent\textit{Rate your response to the following statements from 1 (Strongly disagree) to 7 (Strongly agree).}

\begin{enumerate}
    \item \textbf{I explored the topic in enough depth to develop a well-formed line of inquiry.} \\
    1 (Strongly disagree) -- 2 -- 3 -- 4 -- 5 -- 6 -- 7 (Strongly agree)

    \item \textbf{The line of inquiry I identified felt meaningfully different from the work I encountered.} \\
    1 (Strongly disagree) -- 2 -- 3 -- 4 -- 5 -- 6 -- 7 (Strongly agree)

    \item \textbf{The record or materials produced during this task were useful for developing my line of inquiry.} \\
    1 (Strongly disagree) -- 2 -- 3 -- 4 -- 5 -- 6 -- 7 (Strongly agree)

    \item \textbf{It was easy for me to explore the literature using this workflow.} \\
    1 (Strongly disagree) -- 2 -- 3 -- 4 -- 5 -- 6 -- 7 (Strongly agree)

    \item \textbf{This workflow supported me in reflecting on and expressing what interested me in the literature and why.} \\
    1 (Strongly disagree) -- 2 -- 3 -- 4 -- 5 -- 6 -- 7 (Strongly agree)

    \item \textbf{Completing the task required a great deal of effort.} \\
    1 (Strongly disagree) -- 2 -- 3 -- 4 -- 5 -- 6 -- 7 (Strongly agree)

    \item \textbf{What I produced was worth the effort I had to exert to produce it.} \\
    1 (Strongly disagree) -- 2 -- 3 -- 4 -- 5 -- 6 -- 7 (Strongly agree)

    \item \textbf{I enjoyed using this workflow.} \\
    1 (Strongly disagree) -- 2 -- 3 -- 4 -- 5 -- 6 -- 7 (Strongly agree)

    \item \textbf{I would be happy to use this workflow again for a similar task.} \\
    1 (Strongly disagree) -- 2 -- 3 -- 4 -- 5 -- 6 -- 7 (Strongly agree)
\end{enumerate}

\paragraph{Open-Ended Questions}

For the \textit{Usual Workflow} task:

\begin{enumerate}
    \item \textbf{How did the workflow you used in this task compare with how you would usually explore literature and develop a line of inquiry for your own research?} \\
    \textit{[Open response]}
\end{enumerate}

For the \textit{Wistful} task:

\begin{enumerate}
    \item \textbf{Could you see yourself using this workflow in a real-world setting? Why or why not? If so, for what kinds of tasks might it be useful?} \\
    \textit{[Open response]}
\end{enumerate}

For both tasks:

\begin{enumerate}
    \item \textbf{Were there any aspects of the workflow that were particularly helpful in completing the task? If so, how?} \\
    \textit{[Open response]}

    \item \textbf{Were there any aspects of the workflow that interfered with completing the task? If so, how?} \\
    \textit{[Open response]}

    \item \textbf{Imagine you were to continue developing this project after this task. What would you do next? Would you use or adapt the materials you found or produced during this task? If so, how?} \\
    \textit{[Open response]}

    \item \textbf{Please feel invited to use this space to share any additional thoughts or comments you might have about your experience.} \\
    \textit{[Open response]}
\end{enumerate}

\noindent\textit{After Task 1, participants were instructed to pause here and notify the interviewer before continuing.}

\subsubsection{Senior Researcher Form}

The senior-researcher form followed the same topic, familiarity, and Likert structure, with wording adapted to the mentoring scenario.

\paragraph{Task Reflection}

\noindent\textit{Rate your response to the following statements from 1 (Strongly disagree) to 7 (Strongly agree).}

\begin{enumerate}
    \item \textbf{I explored the topic in enough depth to develop a well-formed line of inquiry.} \\
    1 (Strongly disagree) -- 2 -- 3 -- 4 -- 5 -- 6 -- 7 (Strongly agree)

    \item \textbf{The line of inquiry I identified felt meaningfully different from the work I encountered.} \\
    1 (Strongly disagree) -- 2 -- 3 -- 4 -- 5 -- 6 -- 7 (Strongly agree)

    \item \textbf{The record or materials produced during this task were useful for developing my line of inquiry.} \\
    1 (Strongly disagree) -- 2 -- 3 -- 4 -- 5 -- 6 -- 7 (Strongly agree)

    \item \textbf{It was easy for me to explore the literature using this workflow.} \\
    1 (Strongly disagree) -- 2 -- 3 -- 4 -- 5 -- 6 -- 7 (Strongly agree)

    \item \textbf{This workflow supported me in reflecting on and expressing what interested me in the literature and why.} \\
    1 (Strongly disagree) -- 2 -- 3 -- 4 -- 5 -- 6 -- 7 (Strongly agree)

    \item \textbf{Completing the task required a great deal of effort.} \\
    1 (Strongly disagree) -- 2 -- 3 -- 4 -- 5 -- 6 -- 7 (Strongly agree)

    \item \textbf{What I produced was worth the effort I had to exert to produce it.} \\
    1 (Strongly disagree) -- 2 -- 3 -- 4 -- 5 -- 6 -- 7 (Strongly agree)

    \item \textbf{I enjoyed using this workflow.} \\
    1 (Strongly disagree) -- 2 -- 3 -- 4 -- 5 -- 6 -- 7 (Strongly agree)

    \item \textbf{I would be happy to use this workflow again for a similar task.} \\
    1 (Strongly disagree) -- 2 -- 3 -- 4 -- 5 -- 6 -- 7 (Strongly agree)
\end{enumerate}

\paragraph{Open-Ended Questions}

For the \textit{Usual Workflow} task:

\begin{enumerate}
    \item \textbf{How did the workflow you used in this task compare with how you would usually explore literature and develop a line of inquiry to suggest to a student?} \\
    \textit{[Open response]}
\end{enumerate}

For the \textit{Wistful} task:

\begin{enumerate}
    \item \textbf{Which features of the system, if any, were most or least useful for completing the task? Please explain how they supported or interfered with your workflow.} \\
    \textit{[Open response]}

    \item \textbf{Could you see yourself using this workflow in a real-world setting? Why or why not? If so, for what kinds of tasks might it be useful?} \\
    \textit{[Open response]}
\end{enumerate}

For both tasks:

\begin{enumerate}
    \item \textbf{Were there any aspects of the workflow that were particularly helpful in completing the task? If so, how?} \\
    \textit{[Open response]}

    \item \textbf{Were there any aspects of the workflow that interfered with completing the task? If so, how?} \\
    \textit{[Open response]}

    \item \textbf{Imagine you were to continue helping the student after this task. What would you do next? Would you use or adapt the materials you found or produced during this task? If so, how?} \\
    \textit{[Open response]}

    \item \textbf{Please feel invited to use this space to share any additional thoughts or comments you might have about your experience.} \\
    \textit{[Open response]}
\end{enumerate}

\paragraph{End of Form}

\begin{enumerate}
    \item \textbf{Is there anything else you would like us to know about your experience?} \\
    \textit{[Open response]}
\end{enumerate}

\subsection{\textsc{StitchFlow} Example}

\begin{figure}[H]
    \centering
    \includegraphics[width=0.95\textwidth]{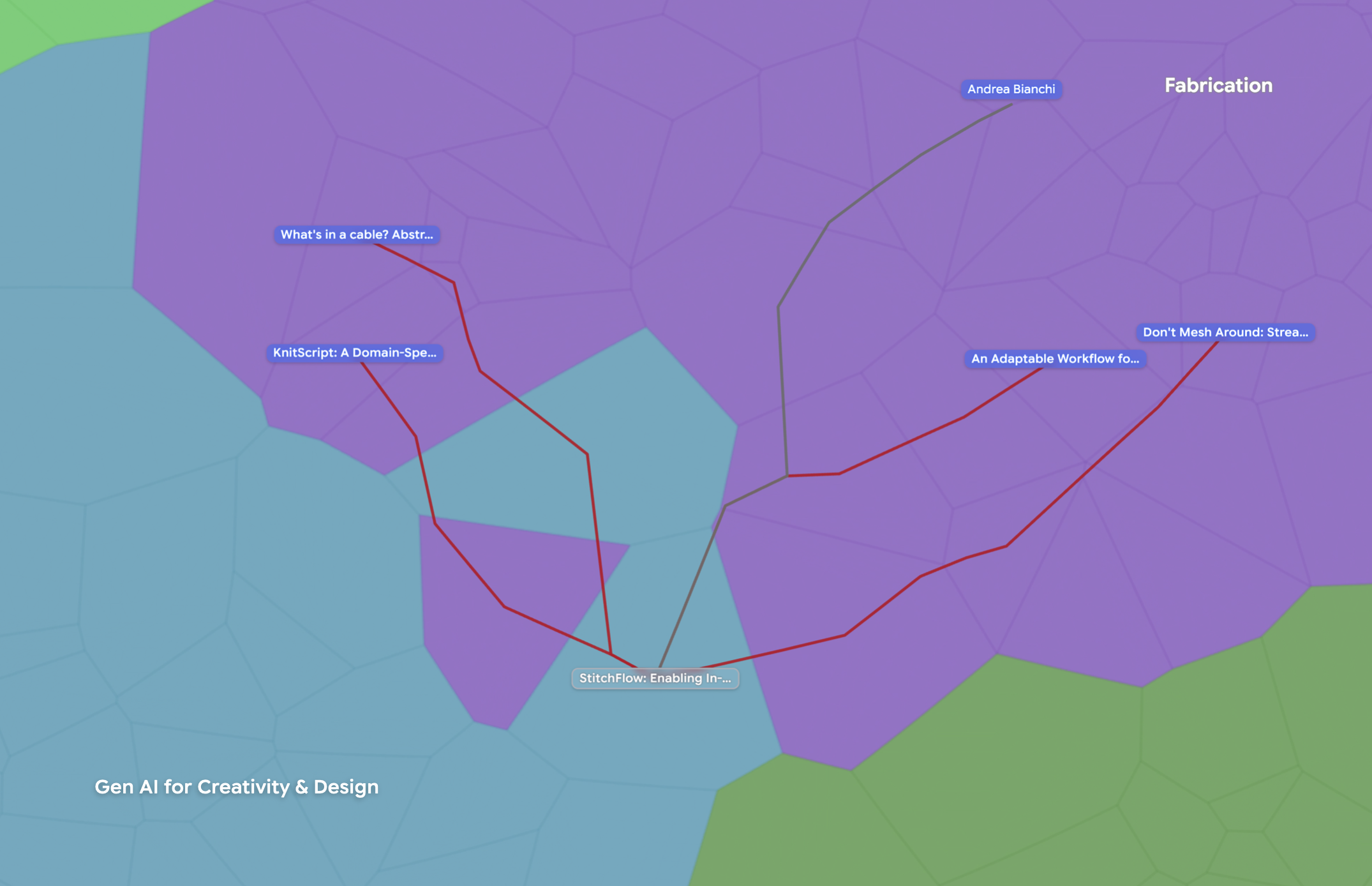}
    \caption{What S7 saw after selecting \textsc{StitchFlow} on the map. Delaunay-routed citation links exposed connections from GenAI Creativity into Fabrication.}
    \Description{TODO}
    \label{fig:stitchflow_map}
\end{figure}

\FloatBarrier

\clearpage
\subsection{All Participants' Spatial Traces}
\label{app:traces}

\begin{figure}[H]
    \centering
    \includegraphics[width=0.95\textwidth]{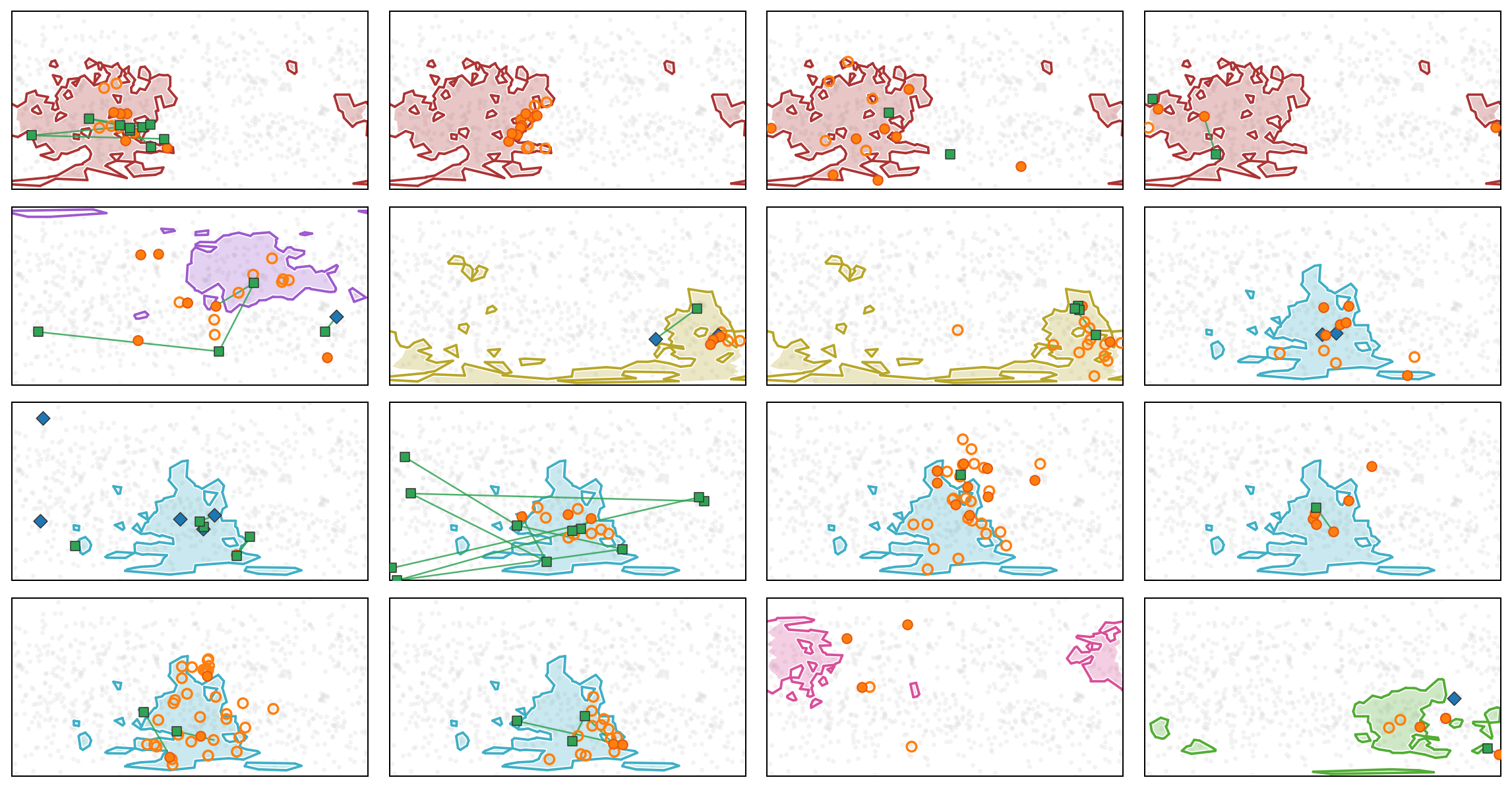}
    \caption{Spatial traces for all 16 participants on the equirectangular map, grouped by assigned continent: AR Interaction \& Authoring (S4, S1, S8, J1); Fabrication (S7); Sensemaking Tools (J8, S2); GenAI for Creativity \& Design (S5, J6, J7, J5, J3, S6, J2); Interaction in VR (J4); and GenAI for Programming (S3).}
    \Description{TODO}
    \label{fig:all_equirect}
\end{figure}

\FloatBarrier

\end{document}